\documentclass[aps,pra,twocolumn,showpacs,preprintnumbers]{revtex4-1}
\usepackage{amsmath}
\usepackage{amssymb}
\usepackage{amsfonts}
\usepackage{graphicx}
\usepackage{dcolumn}
\usepackage{bm}

\begin{document}

\title{New classes of spin chains from $(S\widehat{O}_{(q)}(N)$, $S\widehat{p%
}_{(q)}(N))$ Temperley-Lieb algebras: \\
Data transmission and $(q,N)$ parametrized entanglement entropies}
\author{Amitabha Chakrabarti}
\email{chakra@cpht.polytechnique.fr}
\affiliation{Centre de Physique Th\'eorique, \'{E}cole Polytechnique, 91128 Palaiseau
Cedex, France}
\author{Anirban Chakraborti}
\email{anirban.chakraborti@ecp.fr}
\affiliation{Laboratoire de Math\'{e}matiques Appliqu\'{e}es aux Syst\`{e}mes, \'{E}cole
Centrale Paris, 92290 Ch\^{a}tenay-Malabry, France}
\author{Esteban Guevara Hidalgo}
\email{esteban\_guevarah@yahoo.es}
\affiliation{\'{E}cole Polytechnique, 91128 Palaiseau Cedex, France}
\affiliation{Laboratoire de Math\'ematiques Appliqu\'ees aux Syst\`emes, \'Ecole Centrale
Paris, 92290 Ch\^atenay-Malabry, France}

\begin{abstract}
A Temperley-Lieb algebra is extracted from the operator structure of a new
class of $N^{2}\times N^{2}$ braid matrices presented and studied in
previous papers and designated as $S\widehat{O}_{(q)}(N)$, $S\widehat{p}%
_{(q)}(N)$ for the $q$-deformed orthogonal and symplectic cases
respectively. Spin chain Hamiltonians are derived from such braid matrices
and the corresponding chains are studied. Time evolutions of the chains and
the possibility of transition of data encoded in the parameters of mixed
states from one end to the other are analyzed. The entanglement entropies $%
S(q,N)$ of eigenstates of the crucial operator, namely the $q$-dependent $%
N^{2}\times N^{2}$ projector $P_{0}$ appearing in the corresponding
Hamiltonian are obtained. Study of entanglements generated under the actions
of \ $S\widehat{O}(N)$, $S\widehat{p}(N)$ braid operators, unitarized with
imaginary rapidities is presented as a perspective.
\end{abstract}

\maketitle

\section{Introduction}
\label{intro}
Interesting developments in fault-tolerant quantum computation using the braiding of anyons have brought the theory of braid groups at the very core of topological quantum computing \cite{TQC}. Moreover, the recent study by Kauffman and Lomonaco that the ``Bell matrix'', a specific braiding operator from the solution of the Yang-Baxter equation, is \textit{universal} implies that in principle all quantum gates can be constructed from braiding operators together with single qubit gates \cite{12}.
In another very recent paper, the authors presented a new class of braiding operators from the Temperley-Lieb algebra \cite{TLA} that generalized the Bell matrix to multi-qubit systems, thus unifying the Hadamard and Bell matrices
within the same framework \cite{Ho}. 

Here, we extract the Temperley-Lieb algebra from the operator structure of a new class of $N^{2}\times N^{2}$ braid matrices presented and studied in
previous papers and designated as $S\widehat{O}_{(q)}(N)$, $S\widehat{p}%
_{(q)}(N)$ for the $q$-deformed orthogonal and symplectic cases
respectively \cite{1}. The connection between spin chains and Temperley-Lieb algebras is well-established \cite{5}. We derive the spin chain Hamiltonians from such braid matrices and study the corresponding chains. We then analyze the time evolutions of the chains and
the possibility of transition of data encoded in the parameters of mixed
states from one end to the other, following studies such as Ref. \cite{6}. We further obtain the entanglement entropies $S(q,N)$ of eigenstates of the crucial operator, namely the $q$-dependent $N^{2}\times N^{2}$ projector $P_{0}$ appearing in the corresponding
Hamiltonian. Finally, the study of entanglements generated under the actions
of \ $S\widehat{O}(N)$, $S\widehat{p}(N)$ braid operators, unitarized with
imaginary rapidities, is presented as a perspective.

\section{Braid matrix formalism}
\label{braid}
In the context of exhaustive construction of ``canonically factorized'' forms 
\cite{1} of braid matrices, new classes of solutions with remarkable and intriguing properties were obtained for $SO_{q}(N)$ and $Sp_{q}(N)$. To
distinguish these special cases, they were respectively denoted by $S\widehat{%
O}_{q}(N)$, and $S\widehat{p}_{q}(N)$.
The aspects of that formalism \cite{1} necessary for our present study are
summarized below. 

For $SO_{q}(N)$ and $Sp_{q}(N)$, the standard cases were first expressed \cite{2} in the form%
\begin{equation}
\widehat{R}\left( \theta \right) =\dfrac{f_{+}\left( -\theta \right) }{%
f_{+}\left( \theta \right) }P_{+}+\dfrac{f_{-}\left( -\theta \right) }{%
f_{+}\left( \theta \right) }P_{-}+\dfrac{f_{0}\left( -\theta \right) }{%
f_{0}\left( \theta \right) }P_{0},
\label{1.1}
\end{equation}%
where $(P_{+},P_{-},P_{0})$ form a \textit{complete basis} of projectors, satisfying 
\begin{equation}
P_{i}P_{j}=\delta _{ij}P_{i},\text{ \ \ }P_{+}+P_{-}+P_{0}=I.
  \label{1.2}
\end{equation}%
Since these are $N^{2}\times N^{2}$ $q$-dependent matrices, so is $%
\widehat{R}\left( \theta \right) $. All $\theta $-dependence of $\widehat{R%
}\left( \theta \right) $ is in the coefficients $(f_{+}\left( \theta \right) 
$, $f_{-}\left( \theta \right) $, $f_{0}\left( \theta \right) )$. As
consequence of equation(\ref{1.2})%
\begin{equation}
\widehat{R}\left( \theta \right) \widehat{R}\left( -\theta \right)
=I_{(N^{2}\times N^{2})}=I_{N}\otimes I_{N}\equiv I\otimes I,
  \label{1.3}
\end{equation}%
where $I$ is the ($N\times N$) unit matrix.

The factorization of the coefficients, along with (\ref{1.2}), implies the
``canonical factorization'' of $\widehat{R}\left( \theta \right) :$%
\begin{eqnarray}
\widehat{R}\left( \theta \right) &=&(f_{+}\left( -\theta \right)
P_{+}+f_{-}\left( -\theta \right) P_{-}+f_{0}\left( -\theta \right) P_{0})
\label{1.4} \\
&&\times ((f_{+}\left( \theta \right) )^{-1}P_{+}+(f_{-}\left( \theta
\right) )^{-1}P_{-}+(f_{0}\left( \theta \right) )^{-1}P_{0}).  \notag
\end{eqnarray}%
Like the projectors, the $f$'s will also depend on $q$, the parameter of ``$q$-deformation''.

For $\widehat{R}\left( \theta \right) $to be a braid matrix it must, by
definition, satisfy the braid equation which provides matricial
representation of the third Reidemeister move in the theory of
classification of braids and knots. This means that defining (with
``rapidities'' $\left( \theta ,\theta ^{\prime }\right) $) 
\begin{eqnarray*}
\widehat{R}\left( \theta \right) \otimes I &=&\widehat{R}_{12}\left( \theta
\right) , \\
I\otimes \widehat{R}\left( \theta \right) &=&\widehat{R}_{23}\left( \theta
\right),
\end{eqnarray*}%
we have
\begin{eqnarray}
\widehat{B}_{1} &\equiv &\widehat{R}_{12}\left( \theta \right) \widehat{R}%
_{23}\left( \theta +\theta ^{\prime }\right) \widehat{R}_{12}\left( \theta
^{\prime }\right)  \label{1.5} \\
\widehat{B}_{2} &\equiv &\widehat{R}_{23}\left( \theta ^{\prime }\right) 
\widehat{R}_{12}\left( \theta +\theta ^{\prime }\right) \widehat{R}%
_{23}\left( \theta ^{\prime }\right).  \notag
\end{eqnarray}%
The solutions for the coefficients must be found such that for a given set of
explicitility defined $N^{2}\times N^{2}$ dimensional projectors, one obtains%
\begin{equation}
\widehat{B}_{1}=\widehat{B}_{2}.
  \label{1.6}
\end{equation}

In Ref. \cite{1}, the coefficients for standard known solutions for $SO_{q}(N)$
and $Sp_{q}(N)$ were factorized as in (\ref{1.1}) and new solutions were
obtained, which were studied in subsequent papers \cite{3}.

The \textit{new classes} (denoted as $S\widehat{O}_{q}(N)$ and $S\widehat{p}_{q}(N)$
respectively) correspond to 
\begin{equation}
f_{+}\left( \theta \right) =f_{-}\left( \theta \right) =1  \label{1.7}
\end{equation}%
and hence (with new solutions for $f_{0}$)%
\begin{eqnarray}
\widehat{R}\left( \theta \right) &=&P_{+}+P_{-}+\dfrac{f_{0}\left( -\theta
\right) }{f_{0}\left( \theta \right) }P_{0}  \notag \\
&=&I\otimes I+(\dfrac{f_{0}\left( -\theta \right) }{f_{0}\left( \theta
\right) }-1)P_{0}  \label{1.8} \\
&\equiv &I\otimes I+\omega (\theta )P_{0}.
  \label{1.9}
\end{eqnarray}

In this paper, we analyze for the first time, the properties of $%
P_{0}$ that makes the solution of (\ref{1.9}) possible. Later we consider $%
\omega (\theta )$ in that context. This turns out to be fruitful indeed.

\section{$S\widehat{O}_{q}(N)$, $S\widehat{p}_{q}(N)$ and Temperley-Lieb
Algebra}
\label{special_cases}

We define%
\begin{equation}
(\omega (\theta ),\omega (\theta ^{\prime }),\omega (\theta +\theta ^{\prime
}))\equiv (\omega ,\omega ^{\prime },\omega ^{\prime \prime }).  \label{2.1}
\end{equation}%
Implementing (\ref{1.9}) in (\ref{1.5}), one obtains 
\begin{eqnarray}
(\widehat{B}_{1}-\widehat{B}_{2}) &=&(\omega +\omega ^{\prime }+\omega
\omega ^{\prime }-\omega ^{\prime \prime })((P_{0}\otimes I)-(I\otimes
P_{0}))  \notag \\
&&+\omega \omega ^{\prime }\omega ^{\prime \prime }((P_{0}\otimes
I)(I\otimes P_{0})(P_{0}\otimes I)  \notag \\
&&-(I\otimes P_{0})(P_{0}\otimes I)(I\otimes P_{0})).  \label{2.2}
\end{eqnarray}%

We first define some notations:

\begin{enumerate}
\item[(1)] $\overline{i}\equiv N-i+1$ (when $\overline{\overline{i}}=i$).

\item[(2)] $(ij)$ as the ($N\times N)$ matrix with unity on row $i$ and
column $j$ and all other elements zero.

\item[(3)] The $q$-brackets%
\begin{equation}
\lbrack N\pm 1]=\frac{q^{N\pm 1}-q^{-N\pm 1}}{q-q^{-1}}.  \label{2.3}
\end{equation}
\end{enumerate}

We then have the projectors $P_{0}$ as $N^{2}\times N^{2}$ matrices \cite%
{2}:
\begin{enumerate}
\item[(1)] For $SO_{q}(N),$ ($N=3,4,5,...$)%
\begin{equation}
([N-1]+1)P_{0}=\sum_{i,,j=1}^{N}q^{(\rho \overline{_{i}}-\rho
_{j})}(ij)\otimes (\overline{i}\overline{j}),  \label{2.4}
\end{equation}
and
\item[(2)] For $Sp_{q}(N),$ ($N=2,4,6,...$)%
\begin{equation}
([N+1]-1)P_{0}=\sum_{i,,j=1}^{N}q^{(\rho \overline{_{i}}-\rho
_{j})}(\epsilon _{i}\epsilon _{j})((ij)\otimes (\overline{i}\overline{j})),
\label{2.5}
\end{equation}
where%
\begin{eqnarray}
\epsilon _{i} &=&1,\text{ \ \ }(i\leq N/2)  \notag \\
\epsilon _{i} &=&-1,\text{ \ \ }(i>N/2).  \label{2.6}
\end{eqnarray}%
\end{enumerate}

Note that we restrict $q$ to be real, positive throughout so that one obtains real $P_{0}$. Also, note that the parameters $\rho $ are $N$-tuples:
\begin{enumerate}
\item For $SO(2n+1)$:%
\begin{equation}
\rho :(n-\frac{1}{2},n-\frac{3}{2},...,\frac{1}{2},0,-\frac{1}{2},...,-n+%
\frac{1}{2}).  \label{2.7}
\end{equation}
\item For $SO(2n)$:%
\begin{equation}
\rho :(n-1,n-2,...,1,0,0,-1,...,-n+1).  \label{2.8}
\end{equation}
\item For $Sp(2n):$%
\begin{equation}
\rho :(n,n-1,...,1,-1,...,-n).  \label{2.9}
\end{equation}
\end{enumerate}


The projectors thus defined, can be shown to satisfy%
\begin{eqnarray}
(P_{0}\otimes I)(I\otimes P_{0})(P_{0}\otimes I) &=&k^{-2}(P_{0}\otimes I)
\label{2.10} \\
(I\otimes P_{0})(P_{0}\otimes I)(I\otimes P_{0}) &=&k^{-2}(I\otimes P_{0}),
\label{2.11}
\end{eqnarray}
where%
\begin{eqnarray}
k &=&([N-1]+1),\text{ \ \ for }SO(N)  \label{2.12} \\
k &=&([N+1]-1),\text{ \ \ for }Sp(N).  \label{2.13}
\end{eqnarray}%
This is the core of the Temperley-Lieb algebra to be developed fully for
spin chains, in a following section.

At this stage, implementing (\ref{2.10}) and (\ref{2.11}) in (\ref{2.2}),
one obtains 
\begin{eqnarray}
(\widehat{B}_{1}-\widehat{B}_{2}) &=&(\omega +\omega ^{\prime }+\omega
\omega ^{\prime }-\omega ^{\prime \prime }  \label{2.14} \\
&&+k^{-2}\omega \omega ^{\prime }\omega ^{\prime \prime })((P_{0}\otimes
I)-(I\otimes P_{0})).  \notag
\end{eqnarray}%
Hence the braid equation is satisfied, if 
\begin{equation}
\omega +\omega ^{\prime }+\omega \omega ^{\prime }-\omega ^{\prime \prime
}+k^{-2}\omega \omega ^{\prime }\omega ^{\prime \prime }=0.  \label{2.15}
\end{equation}%
This non-linear functional equation was solved in our previous paper \cite%
{3}. Denoting the special cases as $S\widehat{O}$ and $S\widehat{p}$ henceforward, the solution is given by%
\begin{equation}
\omega (\theta )=(\frac{\sinh (\eta -\theta )}{\sinh (\eta +\theta )}-1),
\label{2.16}
\end{equation}%
where%
\begin{equation}
(e^{\eta }+e^{-\eta })=k=\frac{q^{N-\epsilon }-q^{-N+\epsilon }}{q-q^{-1}}%
+\epsilon  \label{2.17}
\end{equation}%
and 
\begin{eqnarray*}
\epsilon &=&1\text{ \ \ for }S\widehat{O}(N) \\
\epsilon &=&-1\text{ \ \ for }S\widehat{p}(N).
\end{eqnarray*}

We note the following points:

1) Of the three projectors $(P_{+},P_{-},P_{0})$, only one projector, $P_{0}$,
satisfies (\ref{2.10}) and (\ref{2.11}). Instead of (\ref{1.9}), if we start
with 
\begin{equation*}
\widehat{R}\left( \theta \right) =I\otimes I+X_{\pm }(\theta )P_{\pm },
\end{equation*}%
then no solution is obtained, since $P_{\pm }$ do not satisfy the analogues of (\ref%
{2.10}) and (\ref{2.11}). Hence, $P_{0}$ is the crucial operator.

2) An adequate solution of the non-linear functional equation in two
variables ($\theta ,\theta ^{\prime }$) is not evident to start with. But it
does exist and accordingly, an explicit manageable solution (given by (\ref{2.16}) and (\ref%
{2.17})) is obtained.

3) We can re-write (\ref%
{2.17}) as
\begin{equation}
\cosh \eta =\frac{1}{2}k=\frac{1}{2}([N\mp 1]\pm 1)  
\label{3.3}
\end{equation}%
(upper signs for $S\widehat{O}(N)$ and lower signs for $S\widehat{p}(N)$).
 We can also write %
\begin{equation}
\sinh \eta =\pm \frac{1}{2}\sqrt{k^{2}-4}  \label{3.4}
\end{equation}%
for both cases $S\widehat{O}(N)$ and $S\widehat{p}(N)$, and hence $\sinh \eta $ can be chosen to be \textit{positive or negative}, a point which we will revisit.

\section{Spin Chain Hamiltonian and Temperley-Lieb Algebra}
\label{hamiltonian}

The chain Hamiltonian (and also higher order conserved quantities \cite{1})
can be obtained directly as follows. 

We define%
\begin{equation}
\overset{\cdot }{\widehat{R}}(0)=(\frac{d}{d\theta }\widehat{R}\left( \theta
\right) )_{\theta =0},  \label{3.5}
\end{equation}%
and with $\overset{\cdot }{(\widehat{R}}(0))_{l,l+1}$ acting on sites $%
(l,l+1)$, the Hamiltonian for an open chain of length $r$ can be written as 
\begin{equation}
H=\sum_{l=1}^{r-1}I\otimes ...\otimes I\otimes (\overset{\cdot }{\widehat{R}}%
(0))_{l,l+1}\otimes I\otimes ...\otimes I , \label{3.6}
\end{equation}%
where  $I$ is the $N\times N$ unit matrix. 

For a closed chain with
circular boundary conditions, there is an additional term with 
\begin{equation}
(\overset{\cdot }{\widehat{R}}(0))_{r,r+1},\text{ \ \ }(r+1\approx 1).
\label{3.7}
\end{equation}%
For our case 
\begin{equation}
\overset{\cdot }{\widehat{R}}(0)=-(2\coth \eta )P_{0}=\mp (\frac{2k}{%
(k^{2}-4)^{1/2}})P_{0}  \label{3.8}
\end{equation}%
for the upper and lower signs in (\ref{3.4}) respectively.

Hence for open 
$r$-chains, we have
\begin{equation}
H=\mp (\frac{2k}{(k^{2}-4)^{1/2}})(\sum_{l=1}^{r-1}I\otimes ...\otimes
I\otimes (P_{0})_{l,l+1}\otimes I\otimes ...\otimes I).  \label{3.9}
\end{equation}

We define 
\begin{equation}
X_{l}\equiv I\otimes ...\otimes I\otimes (P_{0})_{l,l+1}\otimes I\otimes
...\otimes I.  \label{3.10}
\end{equation}%
Then 
\begin{equation}
H=-(2\coth \eta )(\sum_{l=1}^{r-1}X_{l}),  \label{3.11}
\end{equation}%
where $\eta $ is non-zero, and positive or negative according to the sign
chosen in (\ref{3.4}). 

From (\ref{2.10})-(\ref{2.13}) with $k$ given by (%
\ref{3.3}), 
\begin{eqnarray}
X_{l}X_{l+1}X_{l} &=&k^{-2}X_{l}  \notag \\
X_{l}^{2} &=&X_{l}  \notag \\
X_{l}X_{m} &=&X_{m}X_{l}\text{ \ \ }(|l-m|>1).  \label{3.12}
\end{eqnarray}%
Thus the chain Hamiltonian is obtained as a sum over generators of the
Temperley-Lieb algebra, defined by (\ref{3.12}).

Defining 
\begin{equation}
X_{l}^{\prime }=kX_{l}=(e^{\eta }+e^{-\eta })X_{l}^{\prime },  \label{3.13}
\end{equation}%
one obtains 
\begin{eqnarray}
X_{l}^{\prime }X_{l+1}^{\prime }X_{l}^{\prime } &=&X_{l}^{\prime }  \notag \\
X_{l}^{\prime 2} &=&(e^{\eta }+e^{-\eta })X_{l}^{\prime }  \notag \\
X_{l}^{\prime }X_{m}^{\prime } &=&X_{m}^{\prime }X_{l}^{\prime }\text{ \ \ }%
(|l-m|>1),  \label{3.14}
\end{eqnarray}%
a standard form of Temperley-Lieb algebra.

We note the following points:

\begin{enumerate}
\item As we mentioned before, the link between spin chains and Temperley-Lieb algebras is a
well-studied subject \cite{5}. But here we have more than the defining
relations (\ref{3.12}) or (\ref{3.14}). We have, for all $N$, explicit $%
N^{2}\times N^{2}$ matrix realizations of the generators: (implementing (\ref%
{2.4})-(\ref{2.9}) in (\ref{3.10}) and (\ref{3.13})). This, as will be
displayed below, enables one to construct eigenstates and eigenvalues of
chain Hamiltonians for all $N$.

\item The two signs in (\ref{3.4}) will be seen to correspond to
inversion of the sign of eigenvalues of $H$. They correspond to two
different regimes.
\end{enumerate}

\section{Eigenstates and eigenvalues of $P_{0}$ and action of $H$}
\label{eigen}

We start by presenting the action of the projector $P_{0}$ on product states
and then derive that of the Hamiltonian (\ref{3.9}). The definitions (\ref%
{2.3}), (\ref{2.4}) and the explicit particular cases of Appendix A, imply
that the action of $P_{0}$ on general mixed states selects out specific
linear combinations of states%
\begin{equation*}
\left\vert i\overline{i}\right\rangle =\left\vert i\right\rangle \otimes
\left\vert N-i+1\right\rangle.
\end{equation*}

In terms of 
\begin{equation}
P_{0}^{\prime }=([N-\epsilon ]+\epsilon )P_{0},  \label{4.1}
\end{equation}%
one obtains from (\ref{2.4})-(\ref{2.9}), for\ $S\widehat{O}(N)$:%
\begin{equation}
P_{0}^{\prime }(\sum_{a=1}^{N}x_{a}\left\vert a\right\rangle )\otimes
(\sum_{b=1}^{N}y_{b}\left\vert b\right\rangle )=\sum_{i=1}^{N}q^{(\rho 
\overline{_{i}}-\rho _{i})}x_{i}y\overline{_{i}}\left\vert i\overline{i}%
\right\rangle , \label{4.2}
\end{equation}%
\ and for\ $S\widehat{p}(N)$:%
\begin{equation}
P_{0}^{\prime }(\sum_{a=1}^{N}x_{a}\left\vert a\right\rangle )\otimes
(\sum_{b=1}^{N}y_{b}\left\vert b\right\rangle )=\sum_{i=1}^{N}q^{(\rho 
\overline{_{i}}-\rho _{i})}(\epsilon _{i}x_{i})(\epsilon \overline{_{i}}y%
\overline{_{i}})\left\vert i\overline{i}\right\rangle , \label{4.3}
\end{equation}%
with the $\epsilon$'s defined in (\ref%
{2.6}). 

Explicit results of Appendix A can now be
implemented as follows.

\subsection{$S\widehat{O}(3)$}

\begin{eqnarray}
&&P_{0}^{\prime }(x_{1}\left\vert 1\right\rangle +x_{2}\left\vert
2\right\rangle +x_{\overline{1}}\left\vert \overline{1}\right\rangle ) 
\notag \\
&&\otimes (y_{1}\left\vert 1\right\rangle +y_{2}\left\vert 2\right\rangle
+y_{\overline{1}}\left\vert \overline{1}\right\rangle )  \notag \\
&=&(q^{-1/2}x_{1}y_{\overline{1}}+x_{2}y_{2}+q^{1/2}x\overline{_{1}}%
y_{1})\left\vert \Psi \right\rangle , \label{4.4}
\end{eqnarray}%
where%
\begin{equation}
\left\vert \Psi \right\rangle \equiv (q^{-1/2}\left\vert 1\overline{1}%
\right\rangle +\left\vert 22\right\rangle +q^{1/2}\left\vert \overline{1}%
1\right\rangle ) . \label{4.5}
\end{equation}%
The states $(\left\vert 1\right\rangle ,\left\vert 2\right\rangle
,\left\vert \overline{1}\right\rangle )$ may be taken to correspond to spin
projections $(1,0,-1)$. This $\left\vert \Psi \right\rangle $ turns out to
be an eigenstate of $P_{0}$ with unity for eigenvalue:%
\begin{equation}
P_{0}\left\vert \Psi \right\rangle =\left\vert \Psi \right\rangle \text{ \ \
(from }P_{0}^{\prime }\left\vert \Psi \right\rangle =(q^{-1}+1+q)\left\vert
\Psi \right\rangle \text{)} , \label{4.6}
\end{equation}%
(using (\ref{4.1}) with $N=3$, $([N-1]+1)=(q^{-1}+1+q)$ ). From (\ref{4.4})%
\begin{equation}
P_{0}^{\prime }(\left\vert 1\overline{1}\right\rangle ,\left\vert
22\right\rangle ,\left\vert \overline{1}1\right\rangle
)=(q^{-1/2},1,q^{1/2})\left\vert \Psi \right\rangle  \label{4.7}
\end{equation}%
and 
\begin{equation}
P_{0}^{\prime }\left\vert ij\right\rangle =0\text{ \ \ }(j\neq \overline{i}).
\label{4.8}
\end{equation}%
Strictly analogous results hold for all $S\widehat{O}(N)$ and $S\widehat{p}%
(N).$ This is already pointed out for the cases of Appendix A.

\subsection{$S\widehat{O}(4)$}

\begin{equation*}
P_{0}^{\prime }=(q^{-2}+2+q^{2})P_{0}.
\end{equation*}%
Then
\begin{eqnarray}
&&P_{0}^{\prime }(x_{1}\left\vert 1\right\rangle +x_{2}\left\vert
2\right\rangle +x_{\overline{2}}\left\vert \overline{2}\right\rangle +x_{%
\overline{1}}\left\vert \overline{1}\right\rangle )  \notag \\
&&\otimes (y_{1}\left\vert 1\right\rangle +y_{2}\left\vert 2\right\rangle
+y_{\overline{2}}\left\vert \overline{2}\right\rangle +y_{\overline{1}%
}\left\vert \overline{1}\right\rangle )  \notag \\
&=&(q^{-1}x_{1}y_{\overline{1}}+x_{2}y_{\overline{2}}+x_{\overline{2}%
}y_{2}+qx\overline{_{1}}y_{1})\left\vert \Psi \right\rangle , \label{4.9}
\end{eqnarray}%
where 
\begin{equation}
\left\vert \Psi \right\rangle \equiv q^{-1}\left\vert 1\overline{1}%
\right\rangle +\left\vert 2\overline{2}\right\rangle +\left\vert \overline{2}%
2\right\rangle +q\left\vert \overline{1}1\right\rangle  \label{4.10}
\end{equation}%
and 
\begin{equation}
P_{0}\left\vert \Psi \right\rangle =\left\vert \Psi \right\rangle .
\label{4.11}
\end{equation}%
As in (\ref{4.7}) the non-zero results are obtained for only 
\begin{equation}
P_{0}^{\prime }(\left\vert 1\overline{1}\right\rangle ,\left\vert 2\overline{%
2}\right\rangle ,\left\vert \overline{2}2\right\rangle ,\left\vert \overline{%
1}1\right\rangle )=(q^{-1},1,1,q)\left\vert \Psi \right\rangle . \label{4.12}
\end{equation}%
The states ($\left\vert 1\right\rangle ,\left\vert 2\right\rangle
,\left\vert \overline{2}\right\rangle ,\left\vert \overline{1}\right\rangle $%
) correspond to spin projections ($\frac{3}{2},\frac{1}{2},-\frac{1}{2},-%
\frac{3}{2}$). Our notation here generalizes smoothly to any $N$.

\subsection{$S\widehat{p}(4)$}

\begin{equation}
P_{0}^{\prime }=(q^{-4}+q^{-2}+q^{2}+q^{4})P_{0} . \label{4.13}
\end{equation}%
Then
\begin{eqnarray}
&&P_{0}^{\prime }(x_{1}\left\vert 1\right\rangle +x_{2}\left\vert
2\right\rangle +x_{\overline{2}}\left\vert \overline{2}\right\rangle +x_{%
\overline{1}}\left\vert \overline{1}\right\rangle )  \label{4.14} \\
&&\otimes (y_{1}\left\vert 1\right\rangle +y_{2}\left\vert 2\right\rangle
+y_{\overline{2}}\left\vert \overline{2}\right\rangle +y_{\overline{1}%
}\left\vert \overline{1}\right\rangle )  \notag \\
&=&(q^{-2}x_{1}y_{\overline{1}}+q^{-1}x_{2}y_{\overline{2}}-qx_{\overline{2}%
}y_{2}-q^{2}x\overline{_{1}}y_{1})\left\vert \Psi \right\rangle , \notag
\end{eqnarray}%
where 
\begin{equation}
\left\vert \Psi \right\rangle \equiv q^{-2}\left\vert 1\overline{1}%
\right\rangle +q^{-1}\left\vert 2\overline{2}\right\rangle -q\left\vert 
\overline{2}2\right\rangle -q^{2}\left\vert \overline{1}1\right\rangle
\label{4.15}
\end{equation}%
and
\begin{equation}
P_{0}\left\vert \Psi \right\rangle =\left\vert \Psi \right\rangle .
\label{4.16}
\end{equation}%
Non-zero actions: 
\begin{equation}
P_{0}^{\prime }(\left\vert 1\overline{1}\right\rangle ,\left\vert 2\overline{%
2}\right\rangle ,\left\vert \overline{2}2\right\rangle ,\left\vert \overline{%
1}1\right\rangle )=(q^{-2},q^{-1},-q,-q^{2})\left\vert \Psi \right\rangle
\label{4.17}
\end{equation}%
Let us now consider the action of $H$ 
on an open chain of length $ r $.
%

Similar to (\ref{4.1}), $P_{0}^{\prime }=kP_{0}$, %
we now define
\begin{equation}
X_{l}^{\prime }=kX_{l}=\sum_{l=1}^{r-1}I\otimes ...\otimes I\otimes
(P_{0}^{\prime })_{l,l+1}\otimes I\otimes ...\otimes I\text{ }  \label{4.21}
\end{equation}%
with \ $(l=1,...,r-1)$, 
and
\begin{eqnarray*}
k &=&([N\mp 1]\pm 1) \\
&=&\frac{q^{(N\mp 1)}-q^{-(N\mp 1)}}{q-q^{-1}}\pm 1,
\end{eqnarray*}%
for $S\widehat{O}(N)$ and $S\widehat{p}(N)$ respectively, where we consider
real values of $ q $.

Using (\ref{3.4}), we can now write
\begin{eqnarray}
H &=&\mp \frac{2k}{\sqrt{k^{2}-4}}\sum_{l=1}^{r-1}X_{l}^{\prime }=-\frac{1}{%
\sinh \eta }\sum_{l=1}^{r-1}X_{l}^{\prime }  \label{4.22} \\
&\equiv &-(\sinh \eta )^{-1}H^{\prime } . \label{4.23}
\end{eqnarray}%
Now 
\begin{equation}
H^{\prime }\left\vert i_{1}i_{2}...i_{r}\right\rangle
=\sum_{l=1}^{r-1}\left\vert i_{1}i_{2}...i_{l-1}\right\rangle
((P_{0}^{\prime })\left\vert i_{l}i_{l+1}\right\rangle )\left\vert
i_{l+2}...i_{r}\right\rangle  \label{4.24}
\end{equation}%
and 
\begin{equation}
P_{0}^{\prime }\left\vert i_{l}i_{l+1}\right\rangle =\delta (\overline{i}%
_{l},i_{l+1})q^{(\rho _{\overline{i}}-\rho _{i})}\epsilon _{i}\epsilon _{%
\overline{i}}\left\vert i_{l}\overline{i_{l}}\right\rangle  \label{4.25}
\end{equation}%
with the $n$-tuples defined by (\ref{2.7})-(\ref{2.9}), and with $\epsilon$'s as indicated by (\ref{2.6}) for $S\widehat{p}(N)$ (each $%
\epsilon =1$ for $S\widehat{O}(N)$).

Let us concentrate on the explicit case of $S\widehat{O}(3)$. Consider for example, the $S\widehat{O}(3)$
4-chain with mixed product states 
\begin{eqnarray}
\left\vert X\right\rangle &\equiv &(a_{1}\left\vert 1\right\rangle
+a_{2}\left\vert 2\right\rangle +a_{\overline{1}}\left\vert \overline{1}%
\right\rangle )  \notag \\
&&\otimes (b_{1}\left\vert 1\right\rangle +b_{2}\left\vert 2\right\rangle
+b_{\overline{1}}\left\vert \overline{1}\right\rangle )  \notag \\
&&\otimes (c_{1}\left\vert 1\right\rangle +c_{2}\left\vert 2\right\rangle
+c_{\overline{1}}\left\vert \overline{1}\right\rangle )  \notag \\
&&\otimes (d_{1}\left\vert 1\right\rangle +d_{2}\left\vert 2\right\rangle
+d_{\overline{1}}\left\vert \overline{1}\right\rangle )  \notag \\
&\equiv &\left\vert x\right\rangle _{1}\otimes \left\vert x\right\rangle
_{2}\otimes \left\vert x\right\rangle _{3}\otimes \left\vert x\right\rangle
_{4} . \label{4.26}
\end{eqnarray}%
Defining, as in (\ref{4.5}) 
\begin{equation}
\left\vert \Psi \right\rangle \equiv q^{-1/2}\left\vert 1\overline{1}%
\right\rangle +\left\vert 22\right\rangle +q^{1/2}\left\vert \overline{1}%
1\right\rangle , \label{4.27}
\end{equation}%
we now have
\begin{eqnarray}
H_{4}^{\prime }\left\vert x\right\rangle &=&(a_{1}b_{\overline{1}%
}q^{-1/2}+a_{2}b_{2}+a_{\overline{1}}b_{1}q^{1/2})\left\vert \Psi
\right\rangle \left\vert x\right\rangle _{3}\left\vert x\right\rangle _{4} 
\notag \\
&&+(b_{1}c_{\overline{1}}q^{-1/2}+b_{2}c_{2}+b_{\overline{1}%
}c_{1}q^{1/2})\left\vert x\right\rangle _{1}\left\vert \Psi \right\rangle
\left\vert x\right\rangle _{4}  \notag \\
&&+(c_{1}d_{\overline{1}}q^{-1/2}+c_{2}d_{2}+c_{\overline{1}%
}d_{1}q^{1/2})\left\vert x\right\rangle _{1}\left\vert x\right\rangle
_{2}\left\vert \Psi \right\rangle . \notag \\
&&  \label{4.28}
\end{eqnarray}%
A generalization to a chain of length $r$ is quite straight foward. In
notations that are evident %
\begin{equation}
H^{\prime }(\left\vert x\right\rangle _{1}...\left\vert x\right\rangle
_{r})=\sum_{l=1}^{r-1}f_{l}\left\vert x\right\rangle _{1}...\left\vert
x\right\rangle _{l-1}\left\vert \Psi \right\rangle \left\vert x\right\rangle
_{l+2}...\left\vert x\right\rangle _{r} , \label{4.29}
\end{equation}%
where 
\begin{equation}
f_{l}=(a_{1}^{(l)}b_{\overline{1}%
}^{(l+1)}q^{-1/2}+a_{2}^{(l)}b_{2}^{(l+1)}+a_{\overline{1}%
}^{(l)}b_{1}^{(l+1)}q^{1/2}) . \label{4.30}
\end{equation}%
For $S\widehat{O}(4)$, $S\widehat{p}(4)$ and so on, one can easily implement
(\ref{4.23}) and (\ref{4.24}), with previous definitions.

\section{Time Evolution of Spin Chains and Data Transmission}
\label{time_evol}

\subsection{Evolution in Time}

We are now in a position to start studying the evolution in time $t$ of a
chain under the action of the operator $e^{-iHt}$. As often, we try to
display some basic features by presenting results explicitly for a few
restricted simple cases and indicating how to generalize them.

Consider an $S\widehat{O}(3)$ chain of spins, the projections for spin 1
being denoted as 
\begin{equation}
(\left\vert +\right\rangle ,\left\vert 0\right\rangle ,\left\vert
-\right\rangle )\equiv (\left\vert 1\right\rangle ,\left\vert 0\right\rangle
,\left\vert \overline{1}\right\rangle ) . \label{5.1}
\end{equation}%
From (\ref{4.4})-(\ref{4.8}) one sees then the iterative actions of $H$ on
the buildings blocks 
\begin{equation*}
\left\vert \Psi \right\rangle \left\vert i\right\rangle ,\left\vert
i\right\rangle \left\vert \Psi \right\rangle ,\left\vert i\right\rangle
\left\vert \Psi \right\rangle \left\vert j\right\rangle ,
\end{equation*}%
where 
\begin{equation}
\left\vert \Psi \right\rangle =q^{-1/2}\left\vert 1\overline{1}\right\rangle
+\left\vert 22\right\rangle +q^{1/2}\left\vert \overline{1}1\right\rangle
\label{5.2}
\end{equation}%
and $(i,j)$ take the values $(1,0,\overline{1})$, are the essential
ingredients, along with the basic initial results: 
\begin{eqnarray}
P_{0}^{\prime }(\left\vert 1\overline{1}\right\rangle ,\left\vert
22\right\rangle ,\left\vert \overline{1}1\right\rangle )
&=&(q^{-1}+1+q)P_{0}(\left\vert 1\overline{1}\right\rangle ,\left\vert
22\right\rangle ,\left\vert \overline{1}1\right\rangle )  \notag \\
&=&(q^{-1/2},1,q^{1/2})(q^{-1/2}\left\vert 1\overline{1}\right\rangle  \notag
\\
&&+\left\vert 22\right\rangle +q^{1/2}\left\vert \overline{1}1\right\rangle )
\notag \\
&\equiv &(q^{-1/2},1,q^{1/2})\left\vert \Psi \right\rangle  \notag
\label{5.3}
\end{eqnarray}%
\begin{equation}
P_{0}^{\prime }\left\vert ij\right\rangle =0\text{ \ \ }(j\neq \overline{i}) \notag
\label{5.4}
\end{equation}%
and 
\begin{equation}
kP_{0}\left\vert \Psi \right\rangle \equiv P_{0}^{\prime }\left\vert \Psi
\right\rangle =(q^{-1}+1+q)\left\vert \Psi \right\rangle \equiv k\left\vert
\Psi \right\rangle . \label{5.5}
\end{equation}%
From (\ref{3.4}), (\ref{3.8})-(\ref{3.11}) and (\ref{5.5}), for an open
$ r $-chain 
\begin{equation}
H=\lambda \sum_{l=1}^{r-1}I\otimes ...\otimes I\otimes (P_{0}^{\prime
})_{l,l+1}\otimes I\otimes ...\otimes I , \label{5.6}
\end{equation}%
where (corresponding to the sign of $\eta $ chosen in (\ref{3.4})) 
\begin{equation}
\lambda \equiv \mp \frac{2}{\sqrt{k^{2}-4}} , \label{5.7}
\end{equation}%
which (from (\ref{5.5})) is real for $q$ real, positive (which we assume to
be the case). Note that for a closed chain the summation (\ref{5.6}) would include an
extra term $l=r$ with $(r+1)\approx 1$.

Defining 
\begin{equation}
H\equiv \lambda H^{\prime } , \label{5.8}
\end{equation}%
we consider the series expansion (with $I\otimes I=I_{9}$ for $S\widehat{O}%
(3)$):%
\begin{eqnarray}
e^{-iHt} &=&e^{-i\lambda tH^{\prime }}  \notag \\
&=&I_{9}+(-i\lambda t)H^{\prime }+\frac{1}{2!}(-i\lambda t)^{2}(H^{\prime
})^{2}  \label{5.9} \\
&&+\frac{1}{3!}(-i\lambda t)^{3}(H^{\prime })^{3}+...  \notag
\end{eqnarray}%
up to any chosen order in $t$.

Suppose that the spin states 
\begin{equation*}
\left\vert \Psi \right\rangle \left\vert i\right\rangle ,\left\vert
i\right\rangle \left\vert \Psi \right\rangle ,\left\vert i\right\rangle
\left\vert \Psi \right\rangle \left\vert j\right\rangle
\end{equation*}%
correspond respectively to the sites 
\begin{equation*}
(l,l+1,l+2),\text{ \ \ }(l-1,l,l+1),\text{ \ \ }(l-1,l,l+1,l+2).
\end{equation*}%
Defining 
\begin{eqnarray*}
H_{(3)}^{\prime } &=&P_{0}^{\prime }\otimes I+I\otimes P_{0}^{\prime } \\
H_{(4)}^{\prime } &=&P_{0}^{\prime }\otimes I\otimes I+I\otimes
P_{0}^{\prime }\otimes I+I\otimes I\otimes P_{0}^{\prime }\text{,}
\end{eqnarray*}%
they will always be implicitly assumed to correspond to the appropriate
sites as the relevant parts of the total $H^{\prime }$ acting on the total
chain. Thus $H_{(3)}^{\prime }\left\vert \Psi \right\rangle \left\vert
i\right\rangle $ corresponds to the terms of $H^{\prime }$ acting on the
sites $(l,l+1,l+2)$ and so on.

One obtains from (\ref{4.4})-(\ref{4.8}), 
\begin{eqnarray}
H_{(3)}^{\prime }\left\vert \Psi \right\rangle \left\vert i\right\rangle
&=&k\left\vert \Psi \right\rangle \left\vert i\right\rangle +\left\vert
i\right\rangle \left\vert \Psi \right\rangle  \label{5.10} \\
H_{(3)}^{\prime }\left\vert i\right\rangle \left\vert \Psi \right\rangle
&=&k\left\vert i\right\rangle \left\vert \Psi \right\rangle +\left\vert \Psi
\right\rangle \left\vert i\right\rangle . \label{5.11}
\end{eqnarray}%
Iterating, one obtains (see Appendix B)%
\begin{eqnarray}
(H_{(3)}^{\prime })^{p}(\left\vert \Psi \right\rangle \left\vert
i\right\rangle ) &=&A_{p}\left\vert \Psi \right\rangle \left\vert
i\right\rangle +B_{p}\left\vert i\right\rangle \left\vert \Psi \right\rangle
\label{5.12} \\
(H_{(3)}^{\prime })^{p}\left\vert i\right\rangle \left\vert \Psi
\right\rangle &=&A_{p}\left\vert i\right\rangle \left\vert \Psi
\right\rangle +B_{p}\left\vert \Psi \right\rangle \left\vert i\right\rangle ,
\label{5.13}
\end{eqnarray}%
where%
\begin{eqnarray}
A_{p} &=&\frac{1}{2}((k+1)^{p}+(k-1)^{p})  \label{5.14} \\
B_{p} &=&\frac{1}{2}((k+1)^{p}-(k-1)^{p}) . \label{5.15}
\end{eqnarray}%
Next assuming $j\neq \overline{i}$, we have
\begin{equation}
H_{(4)}^{\prime }(\left\vert i\right\rangle \left\vert \Psi \right\rangle
\left\vert j\right\rangle )=\left\vert \Psi \right\rangle \left\vert
ij\right\rangle +k\left\vert i\right\rangle \left\vert \Psi \right\rangle
\left\vert j\right\rangle +\left\vert ij\right\rangle \left\vert \Psi
\right\rangle  \label{5.16}
\end{equation}%
or 
\begin{equation}
(H_{(4)}^{\prime }-k)(\left\vert i\right\rangle \left\vert \Psi
\right\rangle \left\vert j\right\rangle )=\left\vert ij\right\rangle
\left\vert \Psi \right\rangle +\left\vert \Psi \right\rangle \left\vert
ij\right\rangle . \label{5.17}
\end{equation}%

Now, from (\ref{5.10}), (\ref{5.11}) and (\ref{5.17})%
\begin{equation}
(H_{(4)}^{\prime }-k)^{2}(\left\vert i\right\rangle \left\vert \Psi
\right\rangle \left\vert j\right\rangle )=2\left\vert i\right\rangle
\left\vert \Psi \right\rangle \left\vert j\right\rangle  \label{5.18}
\end{equation}%
since $P_{0}^{\prime }\left\vert ij\right\rangle =0$, $(j\neq \overline{i})$%
. Hence 
\begin{equation}
(H_{(4)}^{\prime }-k)^{2n}(\left\vert i\right\rangle \left\vert \Psi
\right\rangle \left\vert j\right\rangle )=2^{(n-1)}(\left\vert
i\right\rangle \left\vert \Psi \right\rangle \left\vert j\right\rangle )
\label{5.19}
\end{equation}%
\begin{equation}
(H_{(4)}^{\prime }-k)^{2n+1}(\left\vert i\right\rangle \left\vert \Psi
\right\rangle \left\vert j\right\rangle )=2^{(n-1)}(\left\vert
ij\right\rangle \left\vert \Psi \right\rangle +\left\vert \Psi \right\rangle
\left\vert ij\right\rangle . \label{5.20}
\end{equation}%
For $\left\vert j\right\rangle =\left\vert \overline{i}\right\rangle $ there
are additional terms. One obtains%
\begin{equation}
H_{(4)}^{\prime }(\left\vert i\right\rangle \left\vert \Psi \right\rangle
\left\vert \overline{i}\right\rangle )=\left\vert \Psi \right\rangle
\left\vert i\overline{i}\right\rangle +k\left\vert i\right\rangle \left\vert
\Psi \right\rangle \left\vert \overline{i}\right\rangle +\left\vert i%
\overline{i}\right\rangle \left\vert \Psi \right\rangle  \label{5.21}
\end{equation}%
\begin{equation}
H_{(4)}^{\prime }(\left\vert \Psi \right\rangle \left\vert i\overline{i}%
\right\rangle )=k\left\vert \Psi \right\rangle \left\vert i\overline{i}%
\right\rangle +\left\vert i\right\rangle \left\vert \Psi \right\rangle
\left\vert \overline{i}\right\rangle +q^{\delta _{i}}\left\vert \Psi
\right\rangle \left\vert \Psi \right\rangle  \label{5.22}
\end{equation}%
\begin{equation}
H_{(4)}^{\prime }(\left\vert i\overline{i}\right\rangle \left\vert \Psi
\right\rangle )=k\left\vert i\overline{i}\right\rangle \left\vert \Psi
\right\rangle +\left\vert i\right\rangle \left\vert \Psi \right\rangle
\left\vert \overline{i}\right\rangle +q^{\delta _{i}}\left\vert \Psi
\right\rangle \left\vert \Psi \right\rangle , \label{5.23}
\end{equation}%
where $\delta _{i}=(-\frac{1}{2},0,\frac{1}{2})$ respectively for%
\begin{equation}
i=(1,2,\overline{1}) . \label{5.24}
\end{equation}%
Hence,%
\begin{eqnarray}
(H_{(4)}^{\prime })^{2}(\left\vert i\right\rangle \left\vert \Psi
\right\rangle \left\vert \overline{i}\right\rangle ) &=&(k^{2}+2)\left\vert
i\right\rangle \left\vert \Psi \right\rangle \left\vert \overline{i}%
\right\rangle  \notag \\
&&+2k(\left\vert \Psi \right\rangle \left\vert i\overline{i}\right\rangle
+\left\vert i\overline{i}\right\rangle \left\vert \Psi \right\rangle ) 
\notag \\
&&+2q^{\delta _{i}}\left\vert \Psi \right\rangle \left\vert \Psi
\right\rangle  \label{5.25}
\end{eqnarray}%
and again 
\begin{eqnarray}
H_{(4)}^{\prime }(\left\vert \Psi \right\rangle \left\vert \Psi
\right\rangle ) &=&2k\left\vert \Psi \right\rangle \left\vert \Psi
\right\rangle +\sum_{i}q^{\delta _{i}}\left\vert i\right\rangle \left\vert
\Psi \right\rangle \left\vert \overline{i}\right\rangle  \notag \\
&=&2k\left\vert \Psi \right\rangle \left\vert \Psi \right\rangle
+(q^{-1/2}\left\vert 1\right\rangle \left\vert \Psi \right\rangle \left\vert 
\overline{1}\right\rangle  \notag \\
&&+\left\vert 2\right\rangle \left\vert \Psi \right\rangle \left\vert
2\right\rangle +q^{1/2}\left\vert \overline{1}\right\rangle \left\vert \Psi
\right\rangle \left\vert 1\right\rangle ) . \label{5.26}
\end{eqnarray}%

Using the set (\ref{5.21})-(\ref{5.26}) one can now iterate. The way to
proceed and the essential ingredients have been all presented above. We will
not write down the general result for $(H_{(4)}^{\prime })^{n}(\left\vert
i\right\rangle \left\vert \Psi \right\rangle \left\vert \overline{i}%
\right\rangle ).$ In all the examples above there is one feature in common:
The eigenstates $\left\vert \Psi \right\rangle $ of $P_{0}^{\prime }$ appear
under the action of $H^{\prime }$ and move along the chain under iterations.
They move both forward and backward. There can be multiple $\left\vert \Psi
\right\rangle $ depending on the initial state.\ The iterations above are to
be implemented in (\ref{5.9}). Using systematically the results above one
can start to study the evolution of an initial chain configuration.

\subsection{Transmission of Data along a Chain}

For clarity and relative simplicity we start with an open 6-chain of $S%
\widehat{O}(3)$ spin states (\ref{5.1}). The initial configuration is
assumed to be (at $t=0$) 
\begin{eqnarray}
\left\vert X\right\rangle _{(0)} &=&(c_{1}\left\vert \overline{1}%
1\right\rangle +c_{2}\left\vert 1\overline{1}\right\rangle )\left\vert
1111\right\rangle  \label{5.27} \\
&\equiv &c_{1}\left\vert X\right\rangle _{1}+c_{2}\left\vert X\right\rangle
_{2} . \label{5.28}
\end{eqnarray}%
(We do not immediately normalize $\left\vert X\right\rangle $ for convenient
generalization to more parameters, such as, that to start with (\ref{5.36}%
), considered later). \ As will be shown below, time evolution under the action of $e^{-iHt}$
will generate (at time $ t $) a mutually orthogonal set of states including 
\begin{equation}
\left\vert 1111\right\rangle (d_{1}\left\vert \overline{1}1\right\rangle
+d_{2}\left\vert 1\overline{1}\right\rangle ) . \label{5.29}
\end{equation}%
The other states at a finite non-zero $t$ will be (apart from (\ref{5.27}))
sequences 
\begin{eqnarray}
&&(\left\vert X\right\rangle _{1},\left\vert X\right\rangle _{2},\left\vert
\Psi \right\rangle \left\vert 1111\right\rangle ,\left\vert 1\right\rangle
\left\vert \Psi \right\rangle \left\vert 111\right\rangle ,  \notag \\
&&\left\vert 11\right\rangle \left\vert \Psi \right\rangle \left\vert
11\right\rangle ,\left\vert 111\overline{1}11\right\rangle ,\left\vert
111221\right\rangle ,\left\vert 111122\right\rangle ) , \label{5.30}
\end{eqnarray}%
whose coefficients can be obtained (see Appendix C).

With $t$ increasing, the coefficients of the above set, (i.e. (\ref{5.29}) and (%
\ref{5.30})) will continue to change. Otherwise, the set will be stable, no
new basis states of the 6-chain will appear. This is a consequence of the
specific properties of our $H$. It will be shown that it is sufficient to
implement the series development 
\begin{eqnarray}
e^{-iHt} &=&e^{-i(\lambda t)H^{\prime }}  \notag \\
&=&I_{9}-i(\lambda t)H^{\prime }-\frac{1}{2!}(\lambda t)^{2}(H^{\prime
})^{2}+\frac{i}{3!}(\lambda t)^{3}(H^{\prime })^{3}  \notag \\
&&+\frac{1}{4!}(\lambda t)^{4}(H^{\prime })^{4}-\frac{i}{5!}(\lambda
t)^{5}(H^{\prime })^{5}+O(t^{6}) . \label{5.31}
\end{eqnarray}%
Evaluating finally,
\begin{equation*}
(H^{\prime })^{p}(c_{1}\left\vert X\right\rangle _{1}+c_{2}\left\vert
X\right\rangle _{2})
\end{equation*}%
for $p=(0,1,2,3,4,5)$ one already obtains states of the type (\ref{5.29})
(along with others orthogonal to it as given in (\ref{5.30})). Moreover $%
(d_{1},d_{2})$ is obtained explicitly in terms of $(c_{1},c_{2},\lambda ,t)$
where $t$ is given (for any chosen origin) and from (\ref{5.7})
(restricting the values of $q$ for definiteness, to $q=1+\delta ,\delta >0$) 
\begin{equation}
\lambda =\mp \frac{2}{\sqrt{(3)^{2}-4}}=\mp \frac{2}{\sqrt{5}}  \label{5.32}
\end{equation}%
for $S\widehat{O}(3)$ (i.e. for $k=3$). The sign ambiguity in (\ref{5.7})
corresponds to the two possible determinations of $\eta $ (as explained in (%
\ref{3.4})) corresponding to two possible regimes.

Next, one can easily invert the relations and thus extract $(c_{1},c_{2})$
from $(d_{1},d_{2},\lambda ,t)$. Thus the initial mixed state at the left of
the chain can be extracted by precise observation of the specific mixed
state (\ref{5.29}) at the right end of the chain at a finite time $ t $.

In this precise sense, we say that the initial state $(c_{1}\left\vert 
\overline{1}1\right\rangle +c_{2}\left\vert 1\overline{1}\right\rangle )$ at
the left has been transmitted to the right as $(d_{1}\left\vert \overline{1}%
1\right\rangle +d_{2}\left\vert 1\overline{1}\right\rangle )$ where $%
(c_{1},c_{2})$ can be recovered from $(d_{1},d_{2})$.

From the results of Appendix C one obtains 
\begin{eqnarray}
d_{1} &=&c_{1}x_{1}+c_{2}x_{2}  \label{5.33} \\
d_{2} &=&c_{1}y_{1}+c_{2}y_{2} , \label{5.34}
\end{eqnarray}%
where $(x_{1},y_{1})$, $(x_{2},y_{2})$ are given in the appendix. From (\ref%
{C.19}), (\ref{C.21}) one sees (since $\lambda $ and $k$ are known) that the
coefficient of ($t^{3}$) in $x_{2}$ gives directly $c_{2}$ from $d_{1}.$ One
then easily extracts $c_{1}$ also from the coefficients of powers of $t$ in $%
(d_{1},d_{2})$. Thus our goal is attained.

Apart from the development (\ref{5.9}) in powers of $ t $, we can also set 
\begin{equation}
q=1+\delta  \label{5.35}
\end{equation}%
and assuming $\delta $ to be small, use a series development in powers of $%
\delta $ to extract information more readily concerning the initial state
from that at time $t$. Let us illustrate this, very briefly, using a simple
example.

Generalizing (\ref{5.27}) to 
\begin{eqnarray}
\left\vert x\right\rangle _{(0)} &=&(a\left\vert \overline{1}1\right\rangle
+b\left\vert 22\right\rangle +c\left\vert 1\overline{1}\right\rangle
)\left\vert 1111\right\rangle , \label{5.36} \\
H^{\prime }\left\vert x\right\rangle _{(0)}
&=&(q^{-1/2}a+b+q^{1/2}c)\left\vert \Psi \right\rangle \left\vert
1111\right\rangle  \notag \\
&&+cq^{1/2}\left\vert 1\right\rangle \left\vert \Psi \right\rangle
\left\vert 111\right\rangle  \label{5.37} \\
&=&((a+c)-\frac{1}{2}\delta (a-c)+b)\left\vert \Psi \right\rangle \left\vert
1111\right\rangle  \notag \\
&&+c(1+\frac{1}{2}\delta )\left\vert 1\right\rangle \left\vert \Psi
\right\rangle \left\vert 111\right\rangle +O(\delta ^{2}) . \label{5.38}
\end{eqnarray}%
Thus, in the corresponding generalizations of (\ref{5.33}), (\ref{5.34}) one
can separate $(a+c)$ and $(a-c)$ and hence $(a,c)$ by extracting
coefficients of powers of $\delta .$

This can be more helpful for more elaborate initial states. A lesser number
of powers of $t$ will be needed to extract the initial parameters from the
generalizations of $(d_{1},d_{2})$ above.

The special significance of the point $q=1$ will be emphasized at the end of
Sec. \ref{entropy} in the context of entanglement entropy. Here we note that a supplementary series development about $q=1$ can help in another context.
The dependence of our model on the quantum deformation parameter $q$ is
indeed a central feature.

One may compare our results above with the study of ``Quantum
communication through an unmodulated spin chain'' in \cite{6}. There one has
only Pauli matrices and only two possible spin states. But the Hamiltonian
couples all possible pair of sites and static magnetic fields are present.
The action of $e^{-iHt}$ is studied numerically. The specific structure of
our Hamiltonian (not only as here, for $N=3$ but also for $N>3$ through
straightforward generalization) make explicit computations feasible.

To illustrate the above statements we consider, for $S\widehat{O}(3)$, a
particularly simple initial configuration.

Suppose the chain $C$ is given symbolically (with, $1,2,\overline{1},$
corresponding to spin projections $(+1,0,-1)$ respectively) by 
\begin{equation}
C_{0}=(...1111_{(p)}\overline{1}_{(p+1)}\overline{1}\overline{1}\overline{1}%
...)
\end{equation}
with all sites up to, say $p$ in state $1$ and then all sites in state $%
\overline{1}.$

Gathering together the definitions and notations (\ref{4.21})-(\ref{4.24})
in the compact notation%
\begin{eqnarray}
(-iHt) &=&\zeta \lbrack (P_{0(12)}\otimes I\otimes I\otimes ...)+(I\otimes
P_{0(23)}\otimes I\otimes ...)  \notag \\
&&+(I\otimes I\otimes P_{0(34)}\otimes ...)+...] \\
&\equiv &\sum_{p=0}\zeta (H_{p,p+1}) .
\end{eqnarray}%

To start with, only $H_{p,p+1}$ will have a non-zero action on $C$.%
\begin{equation}
HC_{0}=...111(\Psi _{p,p+1})\overline{1}\overline{1}\overline{1}... ,
\end{equation}%
where $\Psi _{p,p+1}$ (given by (\ref{4.5}) ) is%
\begin{equation}
(q^{-1/2}\left\vert 1\overline{1}\right\rangle +\left\vert 22\right\rangle
+q^{1/2}\left\vert \overline{1}1\right\rangle ).
\end{equation}%
In the action of $e^{-iHt}$, the $\Psi $ states then spread out as follows 
\begin{eqnarray}
(-iHt)^{2} &:&  \notag \\
H\Psi _{p,p+1} &\rightarrow &(\Psi _{p-1,p},\Psi _{p,p+1},\Psi _{p+1,p+2})
\end{eqnarray}%
\begin{eqnarray}
(-iHt)^{3} &:&  \notag \\
H\Psi _{p,p} &\rightarrow &(\Psi _{p-2,p-1},\Psi _{p-1,p},\Psi _{p,p+1}) 
\notag \\
H\Psi _{p,p+1} &\rightarrow &(\Psi _{p-1,p},\Psi _{p,p+1},\Psi _{p+1,p+2}) 
\notag \\
H\Psi _{p+1,p+2} &\rightarrow &(\Psi _{p,p+1},\Psi _{p+1,p+2},\Psi
_{p+2,p+3}) .
\end{eqnarray}%
One already sees, schematically, how the $\Psi $ states are generated, move
forward and backward, crossover and acquire coefficients corresponding to
diferent terms 
\begin{equation*}
\frac{1}{n!}(-iHt)^{n}\text{ \ \ }(n=1,2,3) .
\end{equation*}%
Already the multiplicities counting the contributions from different $%
n$'s (with appropiate coefficients for each order) are for 
\begin{eqnarray}
&&(\Psi _{p-2,p-1},\Psi _{p-1,p},\Psi _{p,p+1},\Psi _{p+1,p+2},\Psi
_{p+2,p+3})  \notag \\
&\rightarrow &(1,3,5,3,1) .
\end{eqnarray}%
After, say $r$ steps (i.e. up to order $t^{r}$), the above sequence is 
\begin{equation*}
(...,2r-5,2r-3,2r-1,2r-3,2r-5,...) .
\end{equation*}%
An initial state less simple can give a much more complex pattern. For initial $C$ 
\begin{eqnarray*}
C_{+-} &:&(1\overline{1}1\overline{1}1...\overline{1}1\overline{1}) \\
C_{-+} &:&(\overline{1}1\overline{1}1\overline{1}...1\overline{1}1) \\
C_{00} &:&(22222...222) ,
\end{eqnarray*}%
even the term $(-iHt)$ in the expansion of $e^{-iHt}$\ generates $\Psi $
states for each pair of sites $(p,p+1)$.

These states $\Psi $ are the basic building blocks of our formalism. They
will be seen to be \textit{entangled} states, and their $(q,N)$-dependent entropy will be
studied in the following section.

\section{$(q,N)$-dependent entanglement entropy of eigenstates of $P_{0}$}
\label{entropy}

Acting on the pure product states $\left\vert i\overline{i}\right\rangle $, (%
$\overline{i}=N-i+1$) for both $S\widehat{O}(N)$ and $S\widehat{p}(N)$ the
projector $P_{0}$ creates its eigenstate 
\begin{equation}
P_{0}\left\vert i\overline{i}\right\rangle =\sum_{j=1}^{N}((P_{0})_{(ji),(%
\overline{j}\overline{i})})\left\vert j\overline{j}\right\rangle \approx
\left\vert \Psi \right\rangle , \label{6.1}
\end{equation}%
the matrix elements of $P_{0}$ in (\ref{6.1}), being defined as in (\ref{2.3}) -
(\ref{2.9}).

\textit{Does $P_{0}$ thus generate entanglement?} We give an affirmative answer
below, evaluate the entanglement entropy to quantify it and analyze the $%
(q,N)$-dependence, $ q $ being the parameter of quantum deformation. We
formulate the $q$-dependence, separately for different values of $N$.

\subsection{$S\widehat{O}(3)$}

As usual, we start with $S\widehat{O}(3)$ (see (\ref{4.4})-(\ref{4.8}))
and (\ref{A.1})-(\ref{A.4})) and as essential step, normalize $\left\vert
\Psi \right\rangle $ as below, denoting it now by $\left\vert \Psi
\right\rangle _{(n)}.$ Define 
\begin{equation*}
\left\vert \Psi \right\rangle _{(n)}=(q^{-1}+1+q)^{-1/2}(q^{-1/2}\left\vert
+-\right\rangle +\left\vert 00\right\rangle +q^{1/2}\left\vert
-+\right\rangle )
\end{equation*}%
\begin{equation}
\equiv (q^{-1}+1+q)^{-1/2}(q^{-1/2}\left\vert 1\overline{1}\right\rangle
+\left\vert 22\right\rangle +q^{1/2}\left\vert \overline{1}1\right\rangle )
\label{6.2}
\end{equation}%
That this is an entanglement state, is evident immediately. Attempting to re-express
it as a product state 
\begin{equation*}
(c_{+}\left\vert +\right\rangle +c_{0}\left\vert 0\right\rangle
+c_{-}\left\vert -\right\rangle )(d_{+}\left\vert +\right\rangle
+d_{0}\left\vert 0\right\rangle +d_{-}\left\vert -\right\rangle )
\end{equation*}%
one runs directly into contradictory constraints on $(c_{i},d_{j})$
coefficients.

To quantify the entanglement one first notes that the eigenstate $\left\vert
\Psi \right\rangle _{(n)}$ satisfying 
\begin{equation}
P_{0}\left\vert \Psi \right\rangle _{(n)}=\left\vert \Psi \right\rangle
_{(n)}  \label{6.3}
\end{equation}%
is already Schmidt decomposed as 
\begin{equation}
\left\vert \Psi \right\rangle _{(n)}=\sum_{i}a_{i\overline{i}}\left\vert i%
\overline{i}\right\rangle . \label{6.4}
\end{equation}%
Hence,without passing via the density matrix (and without using a $\log 2$
basis) one obtains the von-Neumann entropy \cite{7, 8} as 
\begin{equation}
S=-\sum_{i}|a_{i\overline{i}}|^{2}\ln |a_{i\overline{i}}|^{2}, \text{ \ \ }%
i=(+,0,-) . \label{6.5}
\end{equation}%
From (\ref{6.2}), (\ref%
{6.4}) and (\ref{6.5}), one obtains 
\begin{eqnarray}
S(q) &=&S(q^{-1})  \label{6.6} \\
&=&\ln (q^{-1}+1+q)-\frac{(q-q^{-1})}{(q^{-1}+1+q)}\ln q  . \notag
\end{eqnarray}%
From $q=1$ one obtains the maximum entropy as 
\begin{equation}
S(1)=S(\max )=\ln 3 . \label{6.7}
\end{equation}%
To first order in $\epsilon >0$, %
\begin{equation}
S(1\mp \epsilon )=\ln 3\pm \frac{2}{3}\epsilon \ln (1\mp \epsilon )<S(1).
\label{6.8}
\end{equation}%
Consistently with (\ref{6.6}) (namely $S(q)=S(q^{-1})$)%
\begin{equation}
S(q)\rightarrow 0\text{ as }q\rightarrow \infty \text{ or }q\rightarrow 0 .
\label{6.9}
\end{equation}%
After displaying the $q$- dependence for $N=3$ we explore below also the $N$-dependence. As a first step we move up from $N=3$ to $N=4$.

\subsection{$S\widehat{O}(4)$}

From (\ref{4.9})-(\ref{4.12}) and (\ref{A.6})-(\ref{A.8}) we define now
the normalized eigenstate of $P_{0}$ as %
\begin{equation}
\left\vert \Psi \right\rangle
_{(n)}=(q^{-2}+2+q^{2})^{-1/2}(q^{-1}\left\vert 1\overline{1}\right\rangle
+\left\vert 2\overline{2}\right\rangle +\left\vert \overline{2}%
2\right\rangle +q\left\vert \overline{1}1\right\rangle ) . \label{6.10}
\end{equation}
The corresponding entropy
is obtained as
\begin{equation}
S(q)=S(q^{-1})=2\ln (q^{-1}+q)-2\frac{(q-q^{-1})}{(q+q^{-1})}\ln q .
\label{6.11}
\end{equation}%
Now, again for $q=1$, 
\begin{eqnarray}
S(\max ) &=&S(1)=\ln 4  \label{6.12} \\
S(1\mp \epsilon ) &=&\ln 4\pm 4\epsilon \ln (1\mp \epsilon )<S(1) .
\label{6.13}
\end{eqnarray}%
Once again (\ref{6.9}) 
\begin{equation}
S(q)\rightarrow 0\text{ as }q\rightarrow \infty \text{ or }q\rightarrow 0 .
\label{6.14}
\end{equation}%
Hence as $N$ increases from $3$ to $4$, $S(\max )$ moves up from $\ln 3$ to $%
\ln 4$ and falls a bit more steeply, but again symmetrically in $(q,q^{-1})$
to vanishing asymptotic values as $q\rightarrow \infty $ and $q\rightarrow 0$

\subsection{$S\widehat{p}(4)$}

For $N=(4,6,8,...)$, i.e., for each such even $N$, one has the projector $%
P_{0}$ for $S\widehat{p}(N)$ as well as for $S\widehat{O}(N)$. Though this
paper is mostly devoted to a detailed study of $S\widehat{O}(3)$, after
showing how the entropy depends on N for $S\widehat{O}(N)$ by presenting the
results for $S\widehat{O}(4)$, we also present briefly the results for $S%
\widehat{p}(4)$ to display both the analogies and the differences in this
respect between $S\widehat{O}(4)$ and $S\widehat{p}(4)$. The relevant
generalization for $N=(6,8,...)$ is straight forward.

For $S\widehat{p}(4)$, starting with (\ref{4.13})-(\ref{4.16}) and (\ref%
{A.9})-(\ref{A.11}) one obtains (as compared to (\ref{6.10}))%
\begin{eqnarray}
\left\vert \Psi \right\rangle _{(n)}
&=&(q^{-4}+q^{-2}+q^{2}+q^{4})^{-1/2}(q^{-2}\left\vert 1\overline{1}%
\right\rangle  \notag \\
&&+q^{-1}\left\vert 2\overline{2}\right\rangle -q\left\vert \overline{2}%
2\right\rangle -q^{2}\left\vert \overline{1}1\right\rangle ) , \label{6.15}
\end{eqnarray}%
the corresponding entanglement entropy\ is 
\begin{eqnarray}
S(q) &=&S(q^{-1})=\ln (q^{-4}+q^{-2}+q^{2}+q^{4})  \notag \\
&&-\frac{4q^{4}+2q^{2}-2q^{-2}-4q^{-4}}{q^{-4}+q^{-2}+q^{2}+q^{4}}\ln q .
\label{6.16}
\end{eqnarray}%
Thus again, as for $S\widehat{O}(4)$,

\begin{eqnarray}
S(1) &=&S(\max )=\ln 4  \label{6.17} \\
S(1\mp \epsilon ) &=&\ln 4\pm 10\epsilon \ln (1\mp \epsilon )  \label{6.18}
\\
S(q) &\rightarrow &0\text{ as }q\rightarrow \infty \text{ or }q\rightarrow 0 .
\label{6.19}
\end{eqnarray}%

\begin{figure}
\includegraphics[scale=0.5]{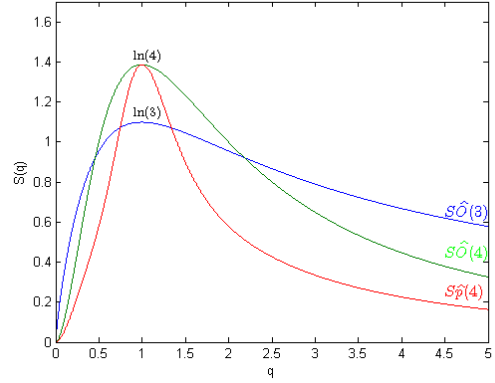}
\caption{$(q,N)$-dependent entanglement entropy.}
\label{fig1}
\end{figure}
The slope, starting from $S(1)$ towards the asymptotic zero values is
steeper as compared to the $S\widehat{O}(4)$ case, as shown in Fig. \ref{fig1}.

One can show (starting with $P_{0}$ defined in (\ref{2.4}), (\ref{2.5}))
that for each $N$, 
\begin{eqnarray}
S(q,N) &=&S(q^{-1},N)  \notag \\
S_{\max }(q,N) &=&S(1,N)=\ln N  \notag \\
S(q,N) &\rightarrow &0\text{ as }q\rightarrow \infty \text{ or }q\rightarrow
0 . \label{6.20}
\end{eqnarray}%
We do not present the explicit (straight forward) computations for general
cases. But one aspect is worth pointing out.

We emphasized in related previous papers \cite{1, 3} that for our special
solutions ($S\widehat{O}(N),$ $S\widehat{p}(N)$) the projector $P_{0}$ and the braid matrix $%
\widehat{R}$ remain non -trivial for $q=1$. This is a remarkable feature
(absent in standard solutions for $SU(N)_{q},$ $SO(N)_{q},$ $Sp(N)_{q}$) as 
was emphasized in Section 3 of reference \cite{1}
and Section 2 of reference \cite{3}.

Now we have found that not only $S\widehat{O}(N)$, $S\widehat{p}(N)$ remain
non trivial for $q=1$, the associated entanglement entropy (for eigenstates
of $P_{0}$) is maximal ($\ln N$) for $q=1$. Thus the striking non-triviality
at $q=1$ acquires further significance.

An adequate study of correlations in presence of multiple states $\left\vert
\Psi \right\rangle $ (as already starting to appear in (\ref{5.25})) will
not be undertaken in this paper. this aspect remains to be explored.

We have shown above that the operators $P_{0}$, acting on any $\left\vert
ij\right\rangle $ either annihilates it (for $j\neq \overline{i}$) or
generates (for $j=\overline{i}$) entangled states $\left\vert \Psi
\right\rangle $ and we have quantified the corresponding entanglement for
all $(q,N)$ by computing the entanglement entropy.

\section{Conclusions and Perspectives}
\label{conclusion}

Starting with the projectors $(P_{+},P_{-},P_{0})$ for $SO_{q}(N)$ and $%
Sp_{q}(N)$ braid matrices and then keeping only $P_{0}$ it was shown how $%
P_{0}$ can generate a Temperley-Lieb algebra and how this property leads
to a remarkable special class of braid matrices (denoted as $S\widehat{O}%
_{q}(N),$ $S\widehat{p}_{q}(N)$) and related spin chains.

Then we have explored certain aspects of such spin chains, using mostly $S%
\widehat{O}_{q}(3)$ examples of chains with free ends to display some
particularly interesting properties.

Time evolution of such chains was studied by evaluating the actions of
successive terms $(-iHt)^{p}$ in the series development of \ $e^{-iHt}$, 
$ H $
being the spin chain Hamiltonian. In particular, we studied in what form the
data encoded in parameters of mixed states at one end of the chain can be
decoded by observing mixed states reaching (as $t$ increases) the other end of
the chain. Most of the relevant computations has been collected together in
the Appendices.

Finally, we have obtained the entanglement entropies $S(q,N)$ of the
eigenstates of $P_{0}$. In particular we obtained $(q,N)$-dependence as%
\begin{eqnarray}
S(q,N) &=&S(q^{-1},N)  \label{7.1} \\
S_{\max }(q,N) &=&S(1,N)=\ln N  \label{7.2} \\
S(q,N) &\rightarrow &0\text{ as }q\rightarrow \infty \text{ or }q\rightarrow
0 . \label{7.3}
\end{eqnarray}%
We pointed out before in Sec. \ref{hamiltonian} that the two possible sign determinations of the
essential parameter $\eta $ correspond to two different regimes for the
energy eigenvalues of the chain Hamiltonian. One may compare and contrast
such a feature with the well-known corresponding ones of the 6-vertex models
(see for example, \cite{9}).

Certain aspects of our classes of spin chains remain to be studied, as
pointed out in Sec. \ref{hamiltonian}.

Another rich perspective is the exploration of various aspects of the braid
matrices we started with (Secs. \ref{braid} and \ref{special_cases}) before extracting from them the chain Hamiltonian (Sec. \ref{hamiltonian}).

In previous papers \cite{10, 11} we studied parametrized entanglements
generated by braid operators rendered unitary by implementing imaginary
rapidities $(i\theta ,i\theta ^{\prime }$) in $\widehat{R}$ matrices of (\ref%
{1.5}). Here again (from (\ref{1.8}) and (\ref{1.9})) 
\begin{eqnarray}
\widehat{R}(i\theta ) &=&P_{+}+P_{-}+\dfrac{f_{0}\left( -i\theta \right) }{%
f_{0}\left( i\theta \right) }P_{0}  \notag \\
&=&I\otimes I+\omega \left( i\theta \right) P_{0}  \label{7.4}
\end{eqnarray}%
can be directly verified to satisfy unitarity, i.e. 
\begin{equation}
(\widehat{R}_{q}(i\theta )^{\dagger }\widehat{R}_{q}(i\theta ))=I_{N}\otimes
I_{N} . \label{7.5}
\end{equation}%
Now one can try to formulate explicitility $(q,N)$-parametrized entanglement
quantifiers of the superpositions of 3-qubit states generated by the action
of the braid operator (see (\ref{1.5}), (\ref{1.6})) on such product states,
as on the l.h.s. of (\ref{4.2}), (\ref{4.3}) generalized to triple products
One can also examine possible teleportation protocols associated to our
class of unitary matrices (see Ref. \cite{12}).

\section{Appendix A}

\subsection*{Explicit $P_{0}$ $(N=3,$ $4)$}

Many basic results of Sec. \ref{eigen} can be read off easily from the matrices $%
P_{0}$ presented below. The matrices $(ij)$ are defined above (\ref{2.3})
as are $(\overline{i}\overline{j})$. The projectors $P_{0}$ are defined by (%
\ref{2.3})-(\ref{2.9}). Their contents for the simplest cases are
displayed below.

\textbf{i) $S\widehat{O}(3)$: \ \ $(N=3;$ $\overline{1}=3,$ $\overline{2}=2) $}

\begin{equation*}
(q^{-1}+1+q)P_{0}\equiv P_{0}^{\prime }
\end{equation*}%
\begin{eqnarray}
&=&q^{-1}(11)\otimes (\overline{1}\overline{1})  \notag \\
&&+q^{-1/2}(12)\otimes (\overline{1}2)+(1\overline{1})\otimes (\overline{1}1)
\notag \\
&&+q^{-1/2}(21)\otimes (2\overline{1})+(22)\otimes (22)  \notag \\
&&+q^{1/2}(2\overline{1})\otimes (21)+(\overline{1}1)\otimes (1\overline{1})
\notag \\
&&+q^{1/2}(\overline{1}2)\otimes (12)+q(\overline{1}\overline{1})\otimes (11)
\notag \\
&&  \label{A.1}
\end{eqnarray}%
\begin{equation}
=%
\begin{vmatrix}
0 & 0 & 0 & 0 & 0 & 0 & 0 & 0 & 0 \\ 
0 & 0 & 0 & 0 & 0 & 0 & 0 & 0 & 0 \\ 
0 & 0 & q^{-1} & 0 & q^{-1/2} & 0 & 1 & 0 & 0 \\ 
0 & 0 & 0 & 0 & 0 & 0 & 0 & 0 & 0 \\ 
0 & 0 & q^{-1/2} & 0 & 1 & 0 & q^{1/2} & 0 & 0 \\ 
0 & 0 & 0 & 0 & 0 & 0 & 0 & 0 & 0 \\ 
0 & 0 & 1 & 0 & q^{1/2} & 0 & q & 0 & 0 \\ 
0 & 0 & 0 & 0 & 0 & 0 & 0 & 0 & 0 \\ 
0 & 0 & 0 & 0 & 0 & 0 & 0 & 0 & 0%
\end{vmatrix}
. \label{A.2}
\end{equation}

Defining the base states 
\begin{equation*}
\left\vert ij\right\rangle \equiv \left\vert i\right\rangle \otimes
\left\vert j\right\rangle ,
\end{equation*}%
the single eigenstate of $P_{0}$ with non-zero eigenvalue is 
\begin{equation}
\left\vert \Psi \right\rangle \equiv q^{-1/2}\left\vert 1\overline{1}%
\right\rangle +\left\vert 22\right\rangle +q^{1/2}\left\vert \overline{1}%
1\right\rangle , \label{A.3}
\end{equation}%
\begin{equation}
P_{0}\left\vert \Psi \right\rangle =\left\vert \Psi \right\rangle .
\label{A.4}
\end{equation}%
All 6 states $\left\vert ij\right\rangle $ with $j\neq i$ are annihilated by 
$P_{0}$. Also, 
\begin{equation}
P_{0}(q^{1/2}\left\vert 1\overline{1}\right\rangle -\left\vert
22\right\rangle )=P_{0}(\left\vert 22\right\rangle -q^{-1/2}\left\vert 
\overline{1}1\right\rangle )=0 . \label{A.5}
\end{equation}%
Corresponding patterns arise for all $N$. They will not be explored in any
detail. 

We present briefly the cases $N=4$.

\textbf{ii) $S\widehat{O}(4)$: \ \ $($ $\overline{1}=4,$ $\overline{2}=3)$}

\begin{equation*}
(q^{-2}+2+q^{2})P_{0}\equiv P_{0}^{\prime }
\end{equation*}%
\begin{eqnarray}
&=&q^{-2}(11)\otimes (\overline{1}\overline{1})+q^{-1}(12)\otimes (\overline{%
1}\overline{2})  \notag \\
&&+q^{-1}(1\overline{2})\otimes (\overline{1}2)+(1\overline{1})\otimes (%
\overline{1}1)  \notag \\
&&+q^{-1}(21)\otimes (\overline{2}\overline{1})+(22)\otimes (\overline{2}%
\overline{2})  \notag \\
&&+(2\overline{2})\otimes (\overline{2}2)+q(2\overline{1})\otimes (\overline{%
2}1)  \notag \\
&&+q^{-1}(2\overline{1})\otimes (2\overline{1})+(\overline{2}2)\otimes (2%
\overline{2})  \notag \\
&&+(\overline{2}\overline{2})\otimes (22)+q(\overline{2}\overline{1})\otimes
(21)  \notag \\
&&+(\overline{1}1)\otimes (1\overline{1})+q(\overline{1}2)\otimes (1%
\overline{2})  \notag \\
&&+q(\overline{1}\overline{2})\otimes (12)+q^{2}(\overline{1}\overline{1}%
)\otimes (11) . \notag \\
&&  \label{A.6}
\end{eqnarray}%
Defining 
\begin{equation}
\left\vert \Psi \right\rangle \equiv q^{-1}\left\vert 1\overline{1}%
\right\rangle +\left\vert 2\overline{2}\right\rangle +\left\vert \overline{2}%
2\right\rangle +q\left\vert \overline{1}1\right\rangle , \label{A.7}
\end{equation}%
\begin{equation}
P_{0}\left\vert \Psi \right\rangle =\left\vert \Psi \right\rangle .
\label{A.8}
\end{equation}

\textbf{iii) $S\widehat{p}(4)$: \ \ $($ $\overline{1}=4,$ $\overline{2}=3)$}

\begin{equation*}
(q^{-4}+q^{-2}+q^{2}+q^{4})P_{0}\equiv P_{0}^{\prime }
\end{equation*}%
\begin{eqnarray}
&=&q^{-4}(11)\otimes (\overline{1}\overline{1})+q^{-3}(12)\otimes (\overline{%
1}\overline{2})  \notag \\
&&-q^{-1}(1\overline{2})\otimes (\overline{1}2)-(1\overline{1})\otimes (%
\overline{1}1)  \notag \\
&&+q^{-3}(21)\otimes (\overline{2}\overline{1})+q^{-2}(22)\otimes (\overline{%
2}\overline{2})  \notag \\
&&-(2\overline{2})\otimes (\overline{2}2)-q(2\overline{1})\otimes (\overline{%
2}1)  \notag \\
&&-q^{-1}(\overline{2}1)\otimes (2\overline{1})-(\overline{2}2)\otimes (2%
\overline{2})  \notag \\
&&+q^{2}(\overline{2}\overline{2})\otimes (22)+q^{3}(\overline{2}\overline{1}%
)\otimes (21)  \notag \\
&&-(\overline{1}1)\otimes (1\overline{1})-q(\overline{1}2)\otimes (1%
\overline{2})  \notag \\
&&+q^{3}(\overline{1}\overline{2})\otimes (12)+q^{4}(\overline{1}\overline{1}%
)\otimes (11) . \notag \\
&&  \label{A.9}
\end{eqnarray}%
Defining 
\begin{equation}
\left\vert \Psi \right\rangle \equiv q^{-2}\left\vert 1\overline{1}%
\right\rangle +q^{-1}\left\vert 2\overline{2}\right\rangle -q\left\vert 
\overline{2}2\right\rangle -q^{2}\left\vert \overline{1}1\right\rangle ,
\label{A.10}
\end{equation}%
\begin{equation}
P_{0}\left\vert \Psi \right\rangle =\left\vert \Psi \right\rangle .
\label{A.11}
\end{equation}%
For $S\widehat{p}(N)$ the blocks with negative signs are anti-diagonally
aligned.

\section{Appendix B}

\subsection*{Iterative Action of $ H $}

As explained in Sec. \ref{time_evol}, in studying $S\widehat{O}(3)$ chains, it is
useful to have ready results for%
\begin{equation*}
H_{(3)}^{m}(\left\vert \Psi \right\rangle \left\vert i\right\rangle
,\left\vert i\right\rangle \left\vert \Psi \right\rangle )
\end{equation*}%
and%
\begin{equation*}
H_{(4)}^{m}(\left\vert i\right\rangle \left\vert \Psi \right\rangle
\left\vert j\right\rangle ) .
\end{equation*}

Here we collect the results indicating the derivations. We consider below $%
H_{(3)}^{\prime }$, $H_{(4)}^{\prime }$ as defined below (\ref{5.9}) in
``notation''. Neccesary multiplicative factors can be easily supplied. We
start with results (\ref{5.10}), (\ref{5.11}). Using them one obtains 
\begin{eqnarray}
H_{(3)}^{\prime }(\left\vert \Psi \right\rangle \left\vert i\right\rangle )
&\equiv &(P_{0}^{\prime }\otimes I+I\otimes P_{0}^{\prime })(\left\vert \Psi
\right\rangle \left\vert i\right\rangle )  \notag \\
&=&k\left\vert \Psi \right\rangle \left\vert i\right\rangle +\left\vert
i\right\rangle \left\vert \Psi \right\rangle , \label{B.1}
\end{eqnarray}%
where $k=(q^{-1}+1+q)$, $i=(1,2,\overline{1})$ also 
\begin{equation}
H_{(3)}^{\prime }(\left\vert i\right\rangle \left\vert \Psi \right\rangle
)=k\left\vert i\right\rangle \left\vert \Psi \right\rangle +\left\vert \Psi
\right\rangle \left\vert i\right\rangle  . \label{B.2}
\end{equation}%
Iterating 
\begin{eqnarray}
(H_{(3)}^{\prime })^{p}(\left\vert \Psi \right\rangle \left\vert
i\right\rangle ) &=&A_{p}\left\vert \Psi \right\rangle \left\vert
i\right\rangle +B_{p}\left\vert i\right\rangle \left\vert \Psi \right\rangle
\label{B.3} \\
(H_{(3)}^{\prime })^{p}(\left\vert i\right\rangle \left\vert \Psi
\right\rangle ) &=&A_{p}\left\vert i\right\rangle \left\vert \Psi
\right\rangle +B_{p}\left\vert \Psi \right\rangle \left\vert i\right\rangle ,
\label{B.4}
\end{eqnarray}%
where 
\begin{eqnarray}
A_{p} &=&\frac{1}{2}((k+1)^{p}+(k-1)^{p})  \label{B.5} \\
B_{p} &=&\frac{1}{2}((k+1)^{p}-(k-1)^{p}) . \label{B.6} 
\end{eqnarray}%
Now 
\begin{eqnarray}
H_{(4)}^{\prime }(\left\vert i\right\rangle \left\vert \Psi \right\rangle
\left\vert j\right\rangle ) &\equiv &(P_{0}^{\prime }\otimes I\otimes
I+I\otimes P_{0}^{\prime }\otimes I  \label{B.7} \\
&&+I\otimes I\otimes P_{0}^{\prime })(\left\vert i\right\rangle \left\vert
\Psi \right\rangle \left\vert j\right\rangle )  \notag \\
&=&\left\vert i\right\rangle \left\vert j\right\rangle \left\vert \Psi
\right\rangle +\left\vert \Psi \right\rangle \left\vert i\right\rangle
\left\vert j\right\rangle  \label{B.8} \\
&&+k\left\vert i\right\rangle \left\vert \Psi \right\rangle \left\vert
j\right\rangle  \notag
\end{eqnarray}%
or 
\begin{equation}
(H_{(4)}^{\prime }-k)(\left\vert i\right\rangle \left\vert \Psi
\right\rangle \left\vert j\right\rangle )=\left\vert i\right\rangle
\left\vert j\right\rangle \left\vert \Psi \right\rangle +\left\vert \Psi
\right\rangle \left\vert i\right\rangle \left\vert j\right\rangle .
\label{B.9}
\end{equation}

For $j\neq \overline{i}$ (when $i\neq \overline{j}$), $P_{0}^{\prime
}\left\vert ij\right\rangle =0$ and from (\ref{B.1}), (\ref{B.2}), (\ref{B.9}%
) 
\begin{equation}
(H_{(4)}^{\prime }-k)^{2}(\left\vert i\right\rangle \left\vert \Psi
\right\rangle \left\vert j\right\rangle )=2\left\vert i\right\rangle
\left\vert \Psi \right\rangle \left\vert j\right\rangle . \label{B.10}
\end{equation}%
Thus 
\begin{eqnarray}
(H_{(4)}^{\prime }-k)^{2n}(\left\vert i\right\rangle \left\vert \Psi
\right\rangle \left\vert j\right\rangle ) &=&2^{n-1}\left\vert
i\right\rangle \left\vert \Psi \right\rangle \left\vert j\right\rangle
\label{B.11} \\
(H_{(4)}^{\prime }-k)^{2n+1}(\left\vert i\right\rangle \left\vert \Psi
\right\rangle \left\vert j\right\rangle ) &=&2^{n-1}(H_{(4)}^{\prime
}-k)\left\vert i\right\rangle \left\vert \Psi \right\rangle \left\vert
j\right\rangle  \notag \\
&=&2^{n-1}(\left\vert i\right\rangle \left\vert j\right\rangle \left\vert
\Psi \right\rangle +\left\vert \Psi \right\rangle \left\vert i\right\rangle
\left\vert j\right\rangle ) . \notag
\end{eqnarray}%
These results can be implemented directly by writing 
\begin{equation*}
e^{-iHt}=e^{-ikt}e^{-i(H-k)t} ,
\end{equation*}%
and using the series development of the last factor. For $j=\overline{i}$
there are extra terms as follows 
\begin{equation}
(H_{(4)}^{\prime }-k)(\left\vert i\right\rangle \left\vert \Psi
\right\rangle \left\vert \overline{i}\right\rangle )=\left\vert i\overline{i}%
\right\rangle \left\vert \Psi \right\rangle +\left\vert \Psi \right\rangle
\left\vert i\overline{i}\right\rangle . \label{B.12}
\end{equation}%
Hence 
\begin{equation}
(H_{(4)}^{\prime }-k)^{2}(\left\vert i\right\rangle \left\vert \Psi
\right\rangle \left\vert \overline{i}\right\rangle )=2\left\vert
i\right\rangle \left\vert \Psi \right\rangle \left\vert \overline{i}%
\right\rangle +2\left\vert \Psi \right\rangle \left\vert \Psi \right\rangle .
\label{B.13}
\end{equation}%
But now 
\begin{eqnarray}
H_{(4)}^{\prime }\left\vert \Psi \right\rangle \left\vert \Psi \right\rangle
&=&2\left\vert \Psi \right\rangle \left\vert \Psi \right\rangle +(I\otimes
P_{0}^{\prime }\otimes I)\left\vert \Psi \right\rangle \left\vert \Psi
\right\rangle  \notag \\
&=&2\left\vert \Psi \right\rangle \left\vert \Psi \right\rangle
+(q^{-1/2}\left\vert 1\right\rangle \left\vert \Psi \right\rangle \left\vert 
\overline{1}\right\rangle  \notag \\
&&+\left\vert 2\right\rangle \left\vert \Psi \right\rangle \left\vert
2\right\rangle +q^{1/2}\left\vert \overline{1}\right\rangle \left\vert \Psi
\right\rangle \left\vert 1\right\rangle ) , \label{B.14}
\end{eqnarray}%

\begin{eqnarray}
&&H_{(4)}^{\prime }(q^{-1/2}\left\vert 1\right\rangle \left\vert \Psi
\right\rangle \left\vert \overline{1}\right\rangle +\left\vert
2\right\rangle \left\vert \Psi \right\rangle \left\vert 2\right\rangle
+q^{1/2}\left\vert \overline{1}\right\rangle \left\vert \Psi \right\rangle
\left\vert 1\right\rangle )  \notag \\
&=&2\left\vert \Psi \right\rangle \left\vert \Psi \right\rangle
+k(q^{-1/2}\left\vert 1\right\rangle \left\vert \Psi \right\rangle
\left\vert \overline{1}\right\rangle  \notag \\
&&+\left\vert 2\right\rangle \left\vert \Psi \right\rangle \left\vert
2\right\rangle +q^{1/2}\left\vert \overline{1}\right\rangle \left\vert \Psi
\right\rangle \left\vert 1\right\rangle ) . \label{B.15}
\end{eqnarray}%
Combining (\ref{B.14}), (\ref{B.15}) one can now iterate.

\section{Appendix C}

\subsection*{Explicit results for a 6-chain}

We present below the iterated action of $H^{\prime }$ (up to fifth order,
namely ($H^{\prime }$)$^{5}$) on the free 6-chain states (for our $S\widehat{%
O}(3)$ case)%
\begin{eqnarray}
\left\vert x\right\rangle _{1} &\equiv &\left\vert \overline{1}%
11111\right\rangle  \label{C.1} \\
\left\vert x\right\rangle _{2} &\equiv &\left\vert 1\overline{1}%
1111\right\rangle . \label{C.2}
\end{eqnarray}%
They will be implemented in Sec. \ref{time_evol} to study, explicitly for a simple
case, the time evolution of our class of spin chains and possible data
transmission with such evolutions.

Here the relevant $H^{\prime }$ is (with $P_{0}^{\prime }$ defined in
Sec. \ref{eigen})%
\begin{equation}
H_{(6)}^{\prime }=\sum_{l=1}^{5}I\otimes I\otimes ...\otimes (P_{0}^{\prime
})_{l,l+1}\otimes ...\otimes I . \label{C.3}
\end{equation}%
The actions of $P_{0}^{\prime }$ on $S\widehat{O}(3)$ states are defined in
Sec. \ref{eigen} and iterative actions are presented, for $H^{\prime }$ acting on $S%
\widehat{O}(3)$ states in Sec. \ref{time_evol}. For the restricted case relevant here
only one needs, for sub-chains of $H_{(6)}^{\prime }$ above%
\begin{eqnarray}
H_{(3)}^{\prime } &\equiv &P_{0}^{\prime }\otimes I+I\otimes P_{0}^{\prime }
\label{C.4} \\
H_{(4)} &\equiv &P_{0}^{\prime }\otimes I\otimes I+I\otimes P_{0}^{\prime
}\otimes I+I\otimes I\otimes P_{0}^{\prime } , \label{C.5}
\end{eqnarray}%
acting respectively on 
\begin{eqnarray}
&&H_{(3)}^{\prime }(\left\vert \Psi \right\rangle \left\vert 1\right\rangle
,\left\vert 1\right\rangle \left\vert \Psi \right\rangle )  \label{C.6} \\
&=&((k\left\vert \Psi \right\rangle \left\vert 1\right\rangle +\left\vert
1\right\rangle \left\vert \Psi \right\rangle ),(\left\vert \Psi
\right\rangle \left\vert 1\right\rangle +k\left\vert 1\right\rangle
\left\vert \Psi \right\rangle ))  \notag \\
H_{(4)}^{\prime }(\left\vert 1\right\rangle \left\vert \Psi \right\rangle
\left\vert 1\right\rangle ) &=&\left\vert \Psi \right\rangle \left\vert
11\right\rangle +k\left\vert 1\right\rangle \left\vert \Psi \right\rangle
\left\vert 1\right\rangle +\left\vert 11\right\rangle \left\vert \Psi
\right\rangle . \label{C.7}
\end{eqnarray}%

Here we have used the basic definitions and results (\ref{4.4})-(\ref{4.8}%
).

Using all these results systematically one obtains the following results in
a straightforward fashion, arranging terms in the order shown below.%
\begin{eqnarray}
H^{\prime }\left\vert x\right\rangle _{1} &=&q^{1/2}\left\vert \Psi
\right\rangle \left\vert 1111\right\rangle  \label{C.8} \\
H^{\prime }\left\vert x\right\rangle _{2} &=&q^{-1}H^{\prime }\left\vert
x\right\rangle _{1}+q^{1/2}\left\vert 1\right\rangle \left\vert \Psi
\right\rangle \left\vert 111\right\rangle , \label{C.9}
\end{eqnarray}%
where $\left\vert \Psi \right\rangle =q^{-1/2}\left\vert 1\overline{1}%
\right\rangle +\left\vert 22\right\rangle +q^{1/2}\left\vert \overline{1}%
1\right\rangle $.%
\begin{eqnarray}
(H^{\prime })^{2}\left\vert x\right\rangle _{1} &=&q^{1/2}(k\left\vert \Psi
\right\rangle \left\vert 1111\right\rangle +\left\vert 1\right\rangle
\left\vert \Psi \right\rangle \left\vert 111\right\rangle )  \label{C.10} \\
(H^{\prime })^{2}\left\vert x\right\rangle _{2} &=&q^{-1}(H^{\prime
})^{2}\left\vert x\right\rangle _{1}+q^{1/2}(\left\vert \Psi \right\rangle
\left\vert 1111\right\rangle  \label{C.11} \\
&&+k\left\vert 1\right\rangle \left\vert \Psi \right\rangle \left\vert
111\right\rangle +\left\vert 11\right\rangle \left\vert \Psi \right\rangle
\left\vert 11\right\rangle ) , \notag
\end{eqnarray}%
where $k=(q^{-1}+1+q)$.%
\begin{eqnarray}
(H^{\prime })^{3}\left\vert x\right\rangle _{1}
&=&q^{1/2}((k^{2}+1)\left\vert \Psi \right\rangle \left\vert
1111\right\rangle  \label{C.12} \\
&&+2k\left\vert 1\right\rangle \left\vert \Psi \right\rangle \left\vert
111\right\rangle +\left\vert 11\right\rangle \left\vert \Psi \right\rangle
\left\vert 11\right\rangle )  \notag \\
(H^{\prime })^{3}\left\vert x\right\rangle _{2} &=&q^{-1}(H^{\prime
})^{3}\left\vert x\right\rangle _{1}+q^{1/2}(2k\left\vert \Psi \right\rangle
\left\vert 1111\right\rangle  \notag \\
&&+(k^{2}+2)\left\vert 1\right\rangle \left\vert \Psi \right\rangle
\left\vert 111\right\rangle +2k\left\vert 11\right\rangle \left\vert \Psi
\right\rangle \left\vert 11\right\rangle  \notag \\
&&+\left\vert 111\right\rangle \left\vert \Psi \right\rangle \left\vert
1\right\rangle ) . \label{C.13}
\end{eqnarray}

\begin{eqnarray}
(H^{\prime })^{4}\left\vert x\right\rangle _{1}
&=&q^{1/2}((k^{3}+3k)\left\vert \Psi \right\rangle \left\vert
1111\right\rangle  \label{C.14} \\
&&+(3k^{2}+2)\left\vert 1\right\rangle \left\vert \Psi \right\rangle
\left\vert 111\right\rangle  \notag \\
&&+3k\left\vert 11\right\rangle \left\vert \Psi \right\rangle \left\vert
11\right\rangle +\left\vert 111\right\rangle \left\vert \Psi \right\rangle
\left\vert 1\right\rangle )  \notag \\
(H^{\prime })^{4}\left\vert x\right\rangle _{2} &=&q^{-1}(H^{\prime
})^{4}\left\vert x\right\rangle _{1}  \label{C.15} \\
&&+q^{1/2}((3k^{2}+2)\left\vert \Psi \right\rangle \left\vert
1111\right\rangle  \notag \\
&&+(k^{3}+6k)\left\vert 1\right\rangle \left\vert \Psi \right\rangle
\left\vert 111\right\rangle  \notag \\
&&+(3k^{2}+3)\left\vert 11\right\rangle \left\vert \Psi \right\rangle
\left\vert 11\right\rangle  \notag \\
&&+4k\left\vert 111\right\rangle \left\vert \Psi \right\rangle \left\vert
1\right\rangle +\left\vert 1111\right\rangle \left\vert \Psi \right\rangle ) .
\notag
\end{eqnarray}

\begin{eqnarray}
(H^{\prime })^{5}\left\vert x\right\rangle _{1}
&=&q^{1/2}((k^{4}+6k^{2}+2)\left\vert \Psi \right\rangle \left\vert
1111\right\rangle  \label{C.16} \\
&&+(4k^{3}+8k)\left\vert 1\right\rangle \left\vert \Psi \right\rangle
\left\vert 111\right\rangle  \notag \\
&&+(6k^{2}+3)\left\vert 11\right\rangle \left\vert \Psi \right\rangle
\left\vert 11\right\rangle  \notag \\
&&+4k\left\vert 111\right\rangle \left\vert \Psi \right\rangle \left\vert
1\right\rangle +\left\vert 1111\right\rangle \left\vert \Psi \right\rangle )
\notag
\end{eqnarray}

\begin{eqnarray}
(H^{\prime })^{5}\left\vert x\right\rangle _{2} &=&q^{-1}(H^{\prime
})^{5}\left\vert x\right\rangle _{1}  \label{C.17} \\
&&+q^{1/2}((4k^{3}+8k)\left\vert \Psi \right\rangle \left\vert
1111\right\rangle  \notag \\
&&+(k^{4}+12k^{2}+5)\left\vert 1\right\rangle \left\vert \Psi \right\rangle
\left\vert 111\right\rangle  \notag \\
&&+(4k^{3}+13k)\left\vert 11\right\rangle \left\vert \Psi \right\rangle
\left\vert 11\right\rangle  \notag \\
&&+(7k^{2}+4)\left\vert 111\right\rangle \left\vert \Psi \right\rangle
\left\vert 1\right\rangle +5k\left\vert 1111\right\rangle \left\vert \Psi
\right\rangle ) . \notag
\end{eqnarray}%
From the preceding results the coefficients (up to $O(t^{5})$) of the states 
$\left\vert 1111\right\rangle $ ($\left\vert \overline{1}1\right\rangle
,\left\vert 22\right\rangle ,\left\vert 1\overline{1}\right\rangle $) are
obtained as given below 
\begin{eqnarray}
e^{-i\lambda tH^{\prime }}\left\vert X\right\rangle _{1} &=&...+\left\vert
1111\right\rangle (x_{1}\left\vert \overline{1}1\right\rangle  \label{C.18}
\\
&&+y_{1}\left\vert 1\overline{1}\right\rangle +z_{1}\left\vert
22\right\rangle )  \notag \\
e^{-i\lambda tH^{\prime }}\left\vert X\right\rangle _{2} &=&...+\left\vert
1111\right\rangle (x_{2}\left\vert \overline{1}1\right\rangle  \label{C.19}
\\
&&+y_{2}\left\vert 1\overline{1}\right\rangle +z_{2}\left\vert
22\right\rangle ) , \notag
\end{eqnarray}%
and 
\begin{eqnarray}
x_{1} &=&\frac{1}{4!}(\lambda t)^{4}-\frac{i}{5!}(\lambda
t)^{5}(4k+q)+O(t^{6})  \label{C.20} \\
y_{1} &=&-\frac{i}{5!}(\lambda t)^{5}+O(t^{6})  \notag \\
z_{1} &=&-\frac{i}{5!}(\lambda t)^{5}q^{-1/2}+O(t^{6})  \notag
\end{eqnarray}%
\begin{eqnarray}
x_{2} &=&\frac{1}{3!}(\lambda t)^{3}+\frac{1}{4!}(\lambda t)^{4}(4k+q+q^{-1})
\label{C.21} \\
&&-\frac{i}{5!}(\lambda t)^{5}(7k^{2}+k(5q+4q^{-1})+5)+O(t^{6})  \notag \\
y_{2} &=&\frac{1}{4!}(\lambda t)^{4}-\frac{i}{5!}(\lambda
t)^{5}(5k+q^{-1})+O(t^{6})  \notag \\
z_{2} &=&-\frac{i}{5!}(\lambda t)^{5}q^{1/2}(5k+q^{-2})+O(t^{6}) . \notag
\end{eqnarray}

\end{document}